\documentclass{aa} 
\usepackage{natbib}
\bibpunct{(}{)}{;}{a}{}{,}

\usepackage{graphicx}
\usepackage{txfonts,textcomp}
\usepackage{nccmath}
\usepackage{hyperref}
\begin{document}

   \title{Thermal Sunyaev-Zeldovich measurements and cosmic infrared background leakage mitigation combining upcoming ground-based telescopes}

   \author{M. Charmetant
          \inst{1}
          \and
          J. Erler\inst{1,2}\fnmsep
          }

   \institute{Argelander-Institut f\"ur Astronomie, Universit\"at Bonn,
              Auf dem H\"ugel 71, 53121 Bonn, Germany\\
              \email{mcharmetant@astro.uni-bonn.de}
         \and
             Deutsches Zentrum f\"ur Luft- und Raumfahrt e.V. (DLR) Projekttr\"ager, Joseph-Beuys-Allee 4, 53113 Bonn, Germany\\
             }

   \date{Received: 31 December 2022/Accepted: 5 June 2023}

% \abstract{}{}{}{}{} 
% 5 {} token are mandatory
 
\abstract
    {The Fred Young Submillimeter Telescope (FYST) and the Simons Observatory Large Aperture Telescope (SO\ LAT) will deliver unprecedented high-resolution measurements of microwave sky emissions. Notably, one of those microwave sky emissions, the thermal Sunyaev-Zeldovich (tSZ) signal, is an essential probe for cluster astrophysics and cosmology. However, an obstacle to its measurement is contamination by the cosmic infrared background (CIB), especially at high frequencies.}
   {Our goal is to assess the detection and purity of tSZ power spectrum measurements from these two telescopes. We demonstrate that FYST's high-frequency coverage helps lower CIB contamination and improves signal detection.}
   {We simulated the various components of the microwave sky at the frequencies, sensitivities, and beam sizes of the upcoming SO LAT and FYST telescopes using full-sky Hierarchical Equal Area isoLatitude Pixelisation (HEALPix) map templates from the Websky simulations and the Python Sky Model (PySM). We used a map-based internal linear combination (ILC) and a constrained ILC (CILC) to extract the tSZ signal and compute residual noises to assess CIB contamination and signal recovery.}
   {We find that the CIB's residual noise power spectrum in the ILC-recovered tSZ is lowered by $\sim 35\%$ on average over the scales $\ell \in [500,5000]$ when SO LAT and FYST are combined compared to when SO LAT is used alone. We find that when using CILC to deproject CIB, the combined abilities of SO LAT and FYST offer a large $\ell \in [1800,3500]$ window in which the recovered tSZ power spectrum is not noise dominated.}
   {}
   {}

   \keywords{galaxies: clusters: general -  methods: data analysis - cosmic microwave background - cosmology: observations - diffuse radiation
               }
   \titlerunning{Thermal Sunyaev-Zeldovich measurements and cosmic infrared background leakage mitigation}
   \authorrunning{Maude Charmetant} 
   \maketitle 
%
%-------------------------------------------------------------------
\section{Introduction}

The cosmic microwave background (CMB) photons we observe have been influenced by their interaction with the baryonic matter as they travel through the Universe. These interactions modify the CMB photons and are called secondary anisotropies of the CMB. The study of these secondary anisotropies allows for the extraction of information about the Large Scale Structure (LSS) of the Universe as well as about the primary anisotropies, the matter content of the Universe. One of the secondary anisotropies, the thermal Sunyaev-Zeldovich (tSZ) effect, is due to the inverse Compton scatting of the CMB photons by the hot electrons in the intracluster medium (ICM) of galaxy clusters. This effect is a crucial tool for cluster detection and was first used by the South Pole Telescope (SPT) \citep{Staniszewski_2009} and the Atacama Cosmology Telescope (ACT) collaborations \citep{Menanteau_2010} and later by the \textit{Planck} collaboration \citep{Planck_clusters} to construct cluster catalogues. The tSZ effect can be used to measure cluster properties, such as temperature \citep{Jens_2018}. It is also crucial for cosmology, as it has a strong dependence on matter clustering $\sigma_{8}$ and matter density $C_{\ell}^{yy} \propto \sigma _{8}^{8.1}\Omega_{\mathrm{m}}^{3.2}$ \citep{Komatsu_2002,Planck_2016d,Bolliet_2018}. By fitting the tSZ power spectrum extracted from data with models of the signal and its contaminants, assuming a Gaussian likelihood function, and applying a Markov chain Monte Carlo (MCMC) analysis with a fixed mass bias $b$, one can constrain these cosmological parameters \citep{Planck_2014_PS}. This implies that a small improvement in the tSZ power spectra statistical significance or a reduction of the bias due to residual contamination from other signals will lead to a significant improvement in measurements of $\sigma _{8}$ and $\Omega_{\mathrm{m}}$. Therefore, a clean, uncontaminated tSZ map is essential. 

The cosmic infrared background (CIB) is the emission from stars and galaxies absorbed and re-emitted by dust in the infrared domain. It is the main contaminant in tSZ reconstruction and is also a tracer of the matter distribution and the LSS \citep{Hurier_2015}. The CIB contamination, or leakage, originates from the spatial correlation between CIB and tSZ and is also due to the component separation techniques,  such as internal linear combination \citep[ILC;][]{Bennett_2003,Eriksen_2004,Park_2007,Kim_2008,Delabrouille_2009}, not being able to perfectly separate the signal from the many other astrophysical emissions present in the sky. Previous studies on CIB leakage in the \textit{Planck} tSZ map made by \cite{Planck_2016d} showed that the amplitude of the leakage increases when the CIB originates from higher redshift populations. The dependence of the leakage on the redshift of the CIB population comes from the fact that the separation methods of components primarily focus on reducing the main contaminant, which at high frequencies is  Galactic thermal dust, and is thus less efficient at reducing the high-z CIB contaminants. This could lead to a bias in tSZ estimation for high-z clusters, as mentioned in \cite{Vikram_2017}. Another study by \cite{Alonso_2018}  showed that the error bars on the tSZ profile are not much improved by the deprojection (i.e. nulling) of the CIB with more constrained component separation methods such as constrained internal linear combination \citep[CILC;][]{Remazeilles_2011}. Dust contamination can reduce the number of detected galaxy clusters by up to $\sim 9\%$ within z $\in[0.3,0.8]$ \citep{Melin_2018}. Similarly, \citep{Zubeldia_2022} demonstrated that CIB contamination in the tSZ signal can lead to $\sim 20\%$ fewer clusters being detected and proposed a publicly available code to produce CIB-free cluster catalogues. 

%, photometric redshift originating from the PSZ2 that were determined using the \texttt{redMaPPer} algorithm \citep{Planck_clusters}.

In the past three decades, several space missions, such as the COsmic Background Explorer \citep[\textit{COBE;}][]{COBE_1994}, the Wilkinson Microwave Anisotropy Probe  \citep[\textit{WMAP;}][]{Bennett_2003}, and \textit{Planck} \citep{Planck_2014},  have observed the sky at microwave frequencies and extracted full-sky maps of the CMB, tSZ, CIB, and other astrophysical components. Upcoming ground-based experiments, such as those using the Simons Observatory Large Aperture Telescope \citep[SO LAT;][]{SO_2019} and the Fred Young Submillimeter Telescope \citep[FYST;][]{CCAT_2021}, will allow for unprecedented measurements of the tSZ and CIB thanks to their high spatial resolution and high sensitivities. 

In this work, our focus is on quantifying the benefits of combining future experiments, such as SO LAT and FYST to reduce the CIB leakage in ILC-reconstructed tSZ maps. A cleaner tSZ map will allow better constraints on the cosmological parameters $\sigma _{8}$ and $\Omega_{\mathrm{m}}$ to be drawn, more accurate cluster counts and detection, and more accurate measurements of the properties of clusters, such as their temperature. Without mitigation, CIB leakage along the line of sight (LoS) could systematically bias measurements of the tSZ effect and lead to higher tSZ measurements, or it could lower such measurements by filling the characteristic tSZ decrement at frequencies below 217 GHz \citep{Schaan_2021}. Obtaining less contaminated signals by combining microwave experiments is common and will be part of the observation strategies of future experiments. For example, previous studies, such as \cite{SO_2019}, have shown that SO LAT will achieve precise measurement of the tSZ power spectrum when combined with \textit{Planck} data. As \textit{Planck} observations are not contaminated by the atmosphere, this combination allows SO LAT to mitigate such contamination on large scales. Similarly, \cite{Melin_2021} constructed a bigger cluster catalogue by combining \textit{Planck} and the South Pole Telescope (SPT-SZ) data. Our work also encourages the combination of all available data from previous and upcoming CMB experiments.

This paper is structured as follows: In Section 2, we formalise the tSZ and CIB description in our study. In Section 3, we describe the modelling of high-resolution maps of the different microwave sky emissions. In particular, we first explain how we generated Galactic foregrounds, the extragalactic components, and their processing. Then, we detail the generation of noises and the handling of beams and masks. This is followed by a conclusion of the output of the sky modelling. In Section 4, we summarise the methods used to extract the tSZ effect, starting with the ILC and followed by CILC. Next, we explain beam convolution handling and the treatment of full-sky maps, and we provide an overview of the methods used. We present a detailed description of our results and an overview in Section 5. Section 6 contains a summary and discussion of the above-mentioned results. 

\section{Formalism}

\subsection{The thermal Sunyaev-Zeldovich effect}

The thermal tSZ effect, predicted by \cite{SZ_1970,SZ_1972}, is the inverse Compton scattering of CMB photons on the hot electrons present in the ICM. A consequence of this effect is the distortion of the CMB blackbody spectrum, due to the shift of the low frequency (corresponding to low energies) CMB photons to higher frequencies (corresponding to higher energies). The temperature change of the CMB due to the tSZ effect can be expressed as: 

\begin{ceqn}
\begin{align}
\label{eqn:1}
    \frac{\Delta T_{\mathrm{CMB}}}{T_{\mathrm{CMB}}} = yf(x_{\nu}) ,
\end{align}
\end{ceqn}

where $T_{\mathrm{CMB}}=2.72548\pm 0.00057$ K \citep{Fixsen_2009} is the temperature of the CMB and $y$ is the dimensionless Compton-y parameter measuring the electron pressure integrated over the LoS proper distance $l$, 

\begin{ceqn}
\begin{align}
    y = \frac{\sigma _{\mathrm{T}}}{m_{\mathrm{e}}c^{2}} \int  P_{\mathrm{e}}(l) \mathrm{d}l,
\end{align}
\end{ceqn}

where $\sigma _{\mathrm{T}}$ is the Thomson cross-section, $m_{\mathrm{e}}$ the electron mass, and $P_{\mathrm{e}}(l)=n_{\mathrm{e}}(l)k_{\mathrm{B}}T_{\mathrm{e}}(l)$ the electron pressure composed of the electron density $n_{\mathrm{e}}$, the Boltzmann constant $k_{\mathrm{B}}$, and the electron temperature $T_{\mathrm{e}}$. In Eq. \eqref{eqn:1}, $f(x_{\nu})$ is the spectral shape of the tSZ \citep{SZ_1972,SZ_1980}: 

\begin{ceqn}
\begin{align}
    f(x_{\nu}) = x_{\nu}\coth \left( \frac{x_{\nu}}{2} \right) -4 = \left ( x_{\nu} \frac{e^{x_{\nu}}+1}{e^{x_{\nu}}-1} - 4 \right ),
\label{eqn:tsz_sed}
\end{align}
\end{ceqn}

where $x_{\nu}=h\nu /k_{\mathrm{B}}T_{\mathrm{CMB}}$, $h$ being the Planck constant and $\nu$ the frequency. The frequency spectrum of the tSZ is characterised by a decrement in the intensity change for frequencies lower than $\nu \sim 217$ GHz and an increment at higher frequencies. To express the frequency variations induced by the tSZ effect in terms of an intensity variation, we use the conversion factor to pass from units of thermodynamic temperature, ($\mathrm{K}_{\mathrm{CMB}}$) \citep{Planck_2014_KCMB} in Eq. \eqref{eqn:1}, to intensity units (Jy/sr),

\begin{ceqn}
\begin{align}
 \Delta I_{\nu} =  \frac{\partial \mathrm{B}_{\nu}(\mathrm{T})}{\partial \mathrm{T}} \Delta \mathrm{T} = \frac{I_{0}}{\mathrm{T}_{\mathrm{CMB}}}h(x_{\nu})\Delta \mathrm{T},
\label{eqn:conv_f}
\end{align}
\end{ceqn}

where $\mathrm{B}_{\nu}(T)$ is the Planck blackbody spectrum at a fixed frequency $\nu$, $I_{0}=2(k_{\mathrm{B}}T_{\mathrm{CMB}})^{3}/(hc)^{2}\approx 270$ $\mathrm{MJysr^{-1}}$, $c$ is the speed of light, and 

\begin{ceqn}
\begin{align}
    h(x_{\nu}) = \frac{x_{\nu}^{4}e^{x_{\nu}}}{(e^{x_{\nu}}-1)^{2}}.
\label{eqn:conv}
\end{align}
\end{ceqn}

Therefore, the tSZ intensity change is given by

\begin{ceqn}
\begin{align}
    \frac{\Delta  I_{\nu}}{I_{0}} = yh(x_{\nu})f(x_{\nu}).
\label{eqn:tsz_MJy}
\end{align}
\end{ceqn}

More details on the Sunyaev-Zeldovich (SZ) effects can be found in the review by \cite{Mroczkowski_2019}. In this study, we place ourselves in a simplistic approach and do not take into account the relativistic correction of the tSZ effect \citep{Wright_1979,Itoh_1998,Pointecouteau_1998,Chluba_2012}, even though \textit{Planck} data has shown that this leads to an underestimation of the amplitude of its power spectrum $C_ {\ell}^{yy}$ \citep{Remazeilles_2018}.

\subsection{The cosmic infrared background}

The CIB is the extragalactic integrated emission produced by all stars and galaxies during their formation and throughout their life. The CIB frequency spectrum peaks in the infrared regime because the majority of sources are far and their light is redshifted by the expansion of the Universe but also because some of this light is absorbed by dust and re-emitted into the infrared regime. This existence of the CIB was predicted by \cite{Partridge_1967} and first detected by  \cite{Puget_1996,Fixsen_1998,Hauser_1998}. Contrary to the tSZ, one frequency spectrum is not sufficient to fully model it. The CIB spectral energy distribution (SED) depends on the redshift of sources.  It can be approximated as a modified blackbody spectrum: 

\begin{ceqn}
\begin{align}
 I_\nu=A_{\mathrm{dust}}\nu ^{\beta }\mathrm{B}_{\nu}(T_{\mathrm{dust}}),
\end{align}
\end{ceqn}

where $A_{\mathrm{dust}}$ is the amplitude of the dust emission, $\beta$ is the dust emissivity spectral index, and $T_{\mathrm{dust}}$ is the temperature of the dust grains. The CIB is a tracer of the entire history of star formation in the Universe. One characteristic of the CIB of particular importance for our study is its non-zero correlation with the tSZ signal. Indeed, some of the unresolved dusty star-forming galaxies constituting the CIB are found in galaxy clusters. Therefore, the spatial correlation between tSZ and CIB is non-zero, and it was measured in \cite{Reichardt_2012,Dunkley_2013,George_2015,Reichardt_2021,Choi_2020}.

\section{Modelling the microwave sky} \label{des_skymodel}

We set up a pipeline called Skymodel to create mock high-resolution maps of the microwave sky that would be observed in future experiments. Our pipeline is based on Hierarchical Equal Area isoLatitude Pixelisation (HEALPix) template maps of the microwave sky from existing simulations as well as the most recent observations of the microwave sky. The microwave sky components are usually separated into two categories: the emissions coming from the Galaxy, called Galactic foregrounds, and the extragalactic components. Our Skymodel deals differently with these two categories, using an existing Python package for the Galactic foregrounds and HEALPix maps from mock simulations for the extragalactic ones.

\subsection{Galactic foregrounds}

Galactic foregrounds are the various emissions originating from the Milky Way and are thus in the foreground of extragalactic emissions. We considered the following Galactic foregrounds: Galactic synchrotron; free-free emission; anomalous microwave emission (AME), which is supposedly produced by spinning dust \citep{AME_2018}; and thermal emission from Galactic dust grains.

\begin{figure}[h!]
    \centering
    \includegraphics[width=0.50\textwidth]{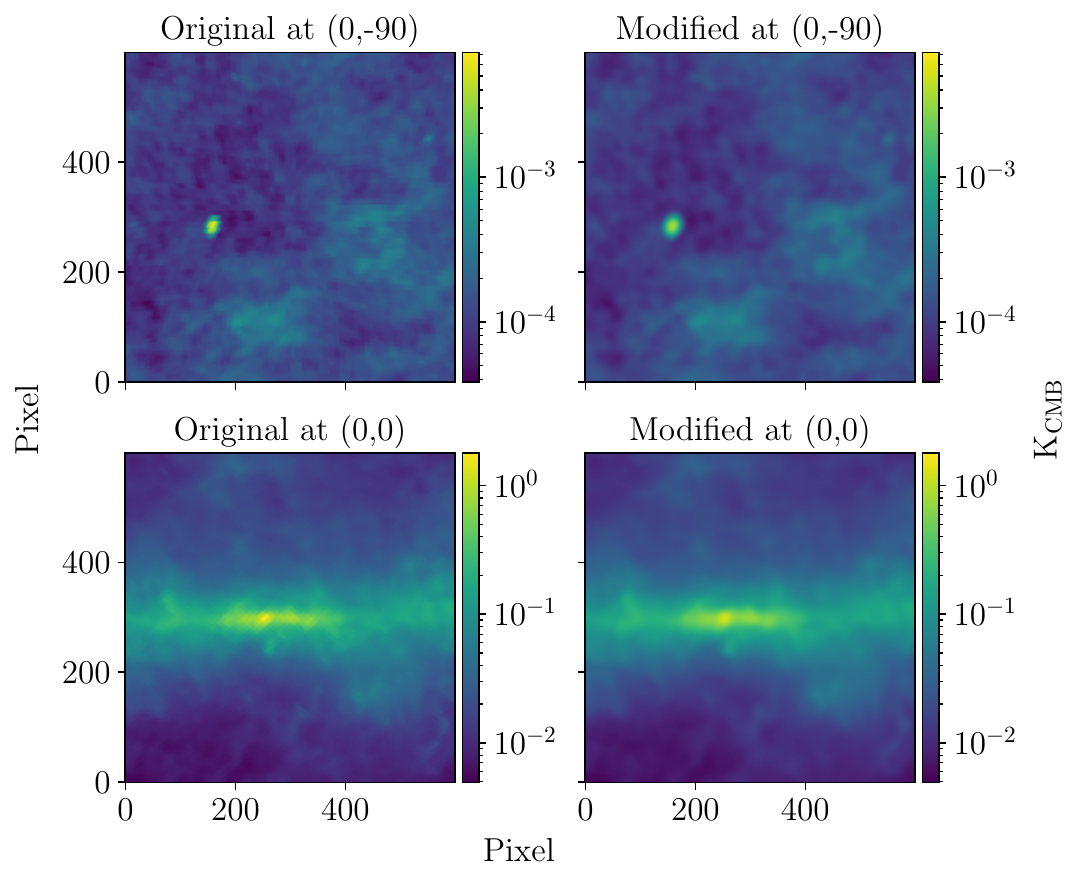}
    \caption{Projections of the PySM sky at $353$GHz composed of free-free, synchrotron, AME, and Galactic dust. The upper row shows cutouts around the Galactic south pole, and the lower row shows cutouts around the Galactic centre. Each cutout is composed of $600\times 600$ pixels of the size $0.86$', making $8.6$° $\times 8.6$° maps. The left column shows cutouts from the original PySM sky, which feature pixelation artefacts due to their pixel size of $6.9$' ($N_\mathrm{side} = 512$). The right column shows modified cutouts obtained by first updating the original map to a higher resolution, $0.86$' ($N_\mathrm{side} = 4096$), by oversampling using the HEALPy function `ud\_grade()' and then  smoothing with a Gaussian kernel of a width that is close the original pixel size of the PySM maps (FWHM = $10$'). This reprocessing of the PySM maps removes the pixelation artefacts while only marginally lowering the spatial resolution.}
    \label{fig:PySM}
\end{figure}

All of the above-mentioned components were generated using the Python Sky Model (PySM) \citep{Ben_2017}. The PySM\footnote{\url{https://github.com/bthorne93/PySM_public}} package is based on Galactic foreground template maps obtained from \cite{Planck_2016_GF} that were extracted using the {\small COMMANDER} algorithm \citep{Eriksen_2004,Eriksen_2008,Planck_2016_COMMANDER,Planck_2020} on \textit{Planck} 2015 raw data, data from \textit{WMAP,} and $408 \, \mathrm{MHz}$ all-sky survey by \cite{Haslam_1981,Haslam_1982}, reprocessed by \cite{Remazeilles_2015}. The original data maps were at a HEALPix \citep{Healpy_2005} pixel resolution of $\mathrm{pix}_{\mathrm{size}}=13.74$' ($N_\mathrm{side} = 256$). The resolution of a HEALPix pixel is defined as the square root of the pixel area. The PySM upgrades them to $6.9$' ($N_\mathrm{side} = 512$) and adds small-scale fluctuations generated randomly by extrapolating the power spectra of those maps \citep{Miville-Deschenes_2007}. We generated each foreground component at the desired frequency using the basic spectral model of PySM (i.e. "s1", "f1", "d1", and "a1"). These maps were then upgraded to a higher $N_\mathrm{side}=4096$ and smoothed with a  $10\arcmin$ full width at half maximum (FWHM) Gaussian to mitigate pixelation artefacts (see Fig. \ref{fig:PySM}).

\subsection{The extragalactic components} \label{sec_extragal}

The extragalactic components include the CMB, tSZ, CIB, and the kinematic Sunyaev-Zeldovich (kSZ) effect. These components were generated at various frequencies using template maps provided by the Websky extragalactic CMB mocks \citep{CITA_2020}, which are HEALPix full-sky maps of the extragalactic sky at $N_\mathrm{side} = 4096$ ($\mathrm{pix}_{\mathrm{size}}=0.86$'). The large-scale structure of the Websky simulations was obtained through a peak-patch and 2nd-order Lagrangian perturbation theory (2LPT) simulation with $12288^3$ particles and a box size of $15.4 \, h^{-1} \, \mathrm{Gpc}$. The authors assumed a flat $\Lambda$ cold dark matter ($\Lambda$CDM) cosmology with $h=0.68$, $\Omega_\mathrm{cdm}=0.261$, $\Omega_{b}=0.049$, $\Omega_{\Lambda}=0.69$, $n_{s}=0.965$, and $\sigma _{8}=0.81$. The simulations can be downloaded from their website.\footnote{\url{https://mocks.cita.utoronto.ca/index.php/Websky_Extragalactic_CMB_Mocks}} An overview of the provided template maps, their frequencies, and units can be seen in Table \ref{tab:skymodel}. 

To generate the tSZ map, \cite{CITA_2020} projected the pressure profiles from hydrodynamical simulations \citep{Battaglia_2012} onto the mass-Peak Patch halo catalogue. They did not consider field contributions, and the minimum halo mass considered to compute their pressure profile was $\sim 1\times 10^{13}M_{\odot}$. Their tSZ power spectrum is in good agreement with the \textit{Planck} data \citep{CITA_2020}. Websky provides all-sky, dimensionless Compton-$y$ maps (linked to the tSZ effect through Eq. \eqref{eqn:1} or \eqref{eqn:tsz_MJy}), which can be multiplied by the tSZ spectrum in Eq. \eqref{eqn:tsz_sed} and \eqref{eqn:conv} to obtain frequency maps of the tSZ. The tSZ signal was simulated by Websky in its non-relativistic limit.

The CIB was generated using the halo occupation distribution model proposed by \cite{Shang_2012} and with the parameters given in \cite{Viero_2013} to fit the CIB power spectrum with the \textit{Herschel} observations. Within this model, the rest-frame SED of a source is a modified blackbody at low frequencies and a power law at high frequencies. The CIB's SED depends on the source redshift, halo mass, and frequency at which it is observed. The Websky halo catalogue is populated with central and satellite galaxies. The field contribution was not considered. The Websky simulations generated CIB maps at the \textit{Planck} High-Frequency Instrument (HFI) frequencies (see Table \ref{tab:skymodel}), and in particular, the $545$ GHz CIB power spectrum was normalised to fit the \textit{Planck} $545$ GHz CIB power spectrum \citep{Planck_2014} at $\ell = 500$. 

\begin{figure}[h!]
    \centering
    \includegraphics[width=0.42\textwidth]{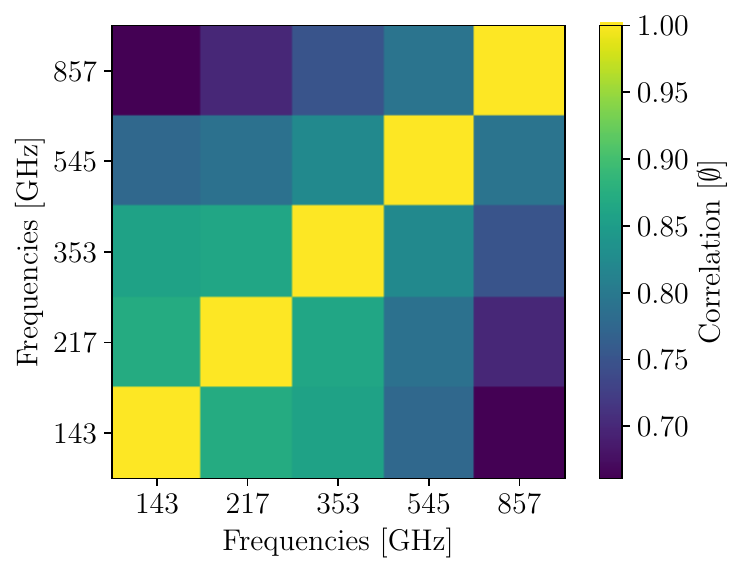}
    \caption{Correlation matrix between the Websky CIB maps provided at the \textit{Planck} HFI frequencies. The axis shows the frequencies of the maps in GHz, and the dimensionless colour bar shows the correlation coefficient.}
    \label{fig:CITA_corr}
\end{figure}

Because of the redshift dependence of the CIB distribution, a decorrelation was expected between the CIB at different frequency channels. The decorrelation was determined in the Websky simulations \citep{CITA_2020} by computing the Pearson product-moment correlation coefficients on the power spectrum of the CIB maps, evaluated at the \textit{Planck} HFI frequency channels. The correlation matrix between the different CIB maps provided by the Websky simulations is illustrated in Fig. \ref{fig:CITA_corr}. The Websky CIB decorrelation values are in agreement with \cite{Planck_2014} and \cite{Lenz_2019}. 

In this work, we extrapolated the CIB model to arbitrary frequencies by fitting each pixel of the existing Websky CIB maps with the following modified blackbody:

\begin{ceqn}
\begin{align}
I_{\nu} = A_{\mathrm{CIB},\nu _{0}} \, \left(\frac{\nu}{\nu_0}\right)^{\beta+3} \frac{e^{\frac{h\nu_0}{k_\mathrm{B}T_{\mathrm{dust}}}}-1}{e^{\frac{h\nu}{k_\mathrm{B}T_{\mathrm{dust}}}}-1},
\label{eqn:mbb_sed}
\end{align}
\end{ceqn}

where $\nu_0 = 353$ GHz. This was done using the Scipy "curve\_fit()" function to derive three all-sky parameter value maps for the variables $\mathrm{A}_{\mathrm{CIB}}$, $T_{\mathrm{dust}}$, and $\beta$. These full-sky maps were injected in Eq. \eqref{eqn:mbb_sed} to produce Websky-like CIB intensity maps at new frequencies.

In the first approximation, we assumed this exact modified black body across frequencies, which is not fully realistic since the CIB is composed of different sources at different redshifts, and their SED is not the same. The Websky simulation includes cross-correlation between the tSZ and CIB, which is in agreement with the \textit{Planck} data results (see Fig. 11 of \cite{CITA_2020}).

For the CMB component, the Websky simulations provided us with two files containing random realisations of the primary T/Q/U $a_{\ell m}$ coefficients, one being unlensed and one being lensed. The lensed file was obtained using the unlensed $a_{\ell m}$ file and the CMB lensing convergence from $0<z<1100$ in the Born approximation. From those $a_{\ell m}$, we could reconstruct a lensed and an unlensed CMB map in $\mu \mathrm{K}_{\mathrm{CMB}}$. Using those maps and the conversion factor (Eq. \eqref{eqn:conv_f}), we could obtain a CMB map at any desired frequency. In the first approximation, we used the unlensed CMB. For the kSZ component, the situation is analogous to the tSZ one. The Websky templates provided frequency-independent maps of the kSZ, which can be seen as a ‘$y_{kSZ}$-map’. We multiplied the maps by the conversion factor in Eq. \eqref{eqn:conv_f} to get a kSZ map at any frequency. The Websky simulations do currently not provide radio point source (RPS) templates, but the Websky team will add them in a future release. This extragalactic component is therefore not accounted for in our study. The different extragalactic full-sky map templates were all converted to the same unit, that is, either temperature units ($\mathrm{K}_{\mathrm{CMB}}$) or intensity units (Jy/sr) units, using Eq. \eqref{eqn:conv_f}.

\begin{table*}[ht]
\begin{center}
\caption{Summary of the base extragalactic templates of the Websky simulations. The templates are generally composed of either one full-sky map or a set of full-sky maps at various frequencies. For the CMB component, the Websky simulations provided a set of $a_{\ell m}$ that are either lensed (L) or unlensed (UL).}
\begin{tabular}{c|c|c|c}
\centering Extragalactic templates from Websky & Type of data & Frequencies (GHz) &  Units \\
\hline
\hline
\rule{0pt}{10pt} \centering Cosmic infrared background (CIB) & HEALPix map &  \{143,217,353,545,857\}
& [MJy/sr] \\
\hline
\rule{0pt}{10pt} \centering  Radio point sources (RPSs) & - & - & - \\ 
\hline
\rule{0pt}{10pt} \centering Cosmic microwave background (CMB) & $\{ a_{\ell m} \}_{L}$ or $\{ a_{\ell m} \}_{UL}$ & - & [$\mu \mathrm{K}_{\mathrm{CMB}}$] 
 \\
\hline
\rule{0pt}{10pt} \centering  thermal Sunyaev-Zeldovich (tSZ) & HEALPix map & Compton-y & [$\emptyset$] \\ 
\hline
\rule{0pt}{10pt} \centering kinematic Sunyaev-Zeldovich (kSZ) & HEALPix map & - & [$\mu \mathrm{K}_{\mathrm{CMB}}$] \\
\label{tab:skymodel}
\end{tabular}
\end{center}
\end{table*}

\subsection{Future experiments}

We simulated the microwave sky as will be seen by two future ground-based telescopes.
One of these telescopes is the FYST of the Cerro Chajnantor Atacama Telescope prime (CCAT-prime) collaboration \citep{CCAT_2021}. It is an upcoming 6 m diameter telescope that will be located at 5600 m on the mountain of Cerro Chajnantor in Chile. Due to its very low precipitable water vapour, this exceptional site will provide one of the best atmospheric transmissions for ground-based microwave observations to date \citep{Bustos_2014}. The crossed-Dragone optical design proposed by \citep{Niemack_2016} will deliver a high-throughput, wide-FoV telescope capable of rapidly scanning large areas of the sky. The first light is expected in 2024.

The second telescope of interest will be part of the Simons Observatory \citep[SO;][]{SO_2019}. The SO will be composed of three refracting 0.4 m small aperture telescopes (SATs) and one cross-Dragone 6 m large aperture telescope (LAT). The four instruments will be located at 5300 m on Cerro Toco in the Atacama Desert in Chile. 

We studied what benefits may result from the combination of FYST with the SO LAT. To do so, we used PySM and the Websky mock sky to simulate the microwave sky at frequencies and sensitivities, meaning the smallest signal a telescope can detect above the random background noise, as well as the beam sizes of future observations conducted with these telescopes, see Table \ref{tab:fsb}.

\begin{table*}[h!]
    \centering
     \caption{Frequencies, beam sizes in FWHM, and baseline instrumental sensitivities for SO LAT and FYST.}
    \begin{tabular}{cc|cc|cc}
    & & \multicolumn{2}{c|}{SO LAT} & \multicolumn{2}{c}{FYST} \\
    Frequency & FWHM & \multicolumn{2}{c|}{Sensitivity} & \multicolumn{2}{c}{Sensitivity} \\
    (GHz) & (arcmin) & ($\mu K_{CMB}$-arcmin) & ($\mu K_{RJ}$-arcmin) & ($\mu K_{CMB}$-arcmin) & ($\mu K_{RJ}$-arcmin) \\
    \hline
    \hline
      93 & 2.2 & 8 & 6.42 & -- & --  \\
      145 & 1.4 & 10 & 5.96 & -- & -- \\
      220 & 1.0 &  -- &  -- & 15 &  4.86 \\
     225 & 1.0 & 22 & 6.80 & -- & -- \\
     279 & 0.9 & 54 & 9.68 & -- & -- \\
      280 & 0.9 & -- & -- & 27 & 4.79 \\
      350 & 0.6 & -- & -- & 105 & 8.38 \\
     405 & 0.5 & -- & -- & 372 & 15 \\
     860 & 0.2 & -- & -- & $5.7\times 10 ^{5}$ & 33.93 \\
    \end{tabular}
    \label{tab:fsb}
\end{table*}

\subsection{Noises, masks, and beams}

\begin{figure}[h!]
    \centering
    \includegraphics[width=0.48\textwidth]{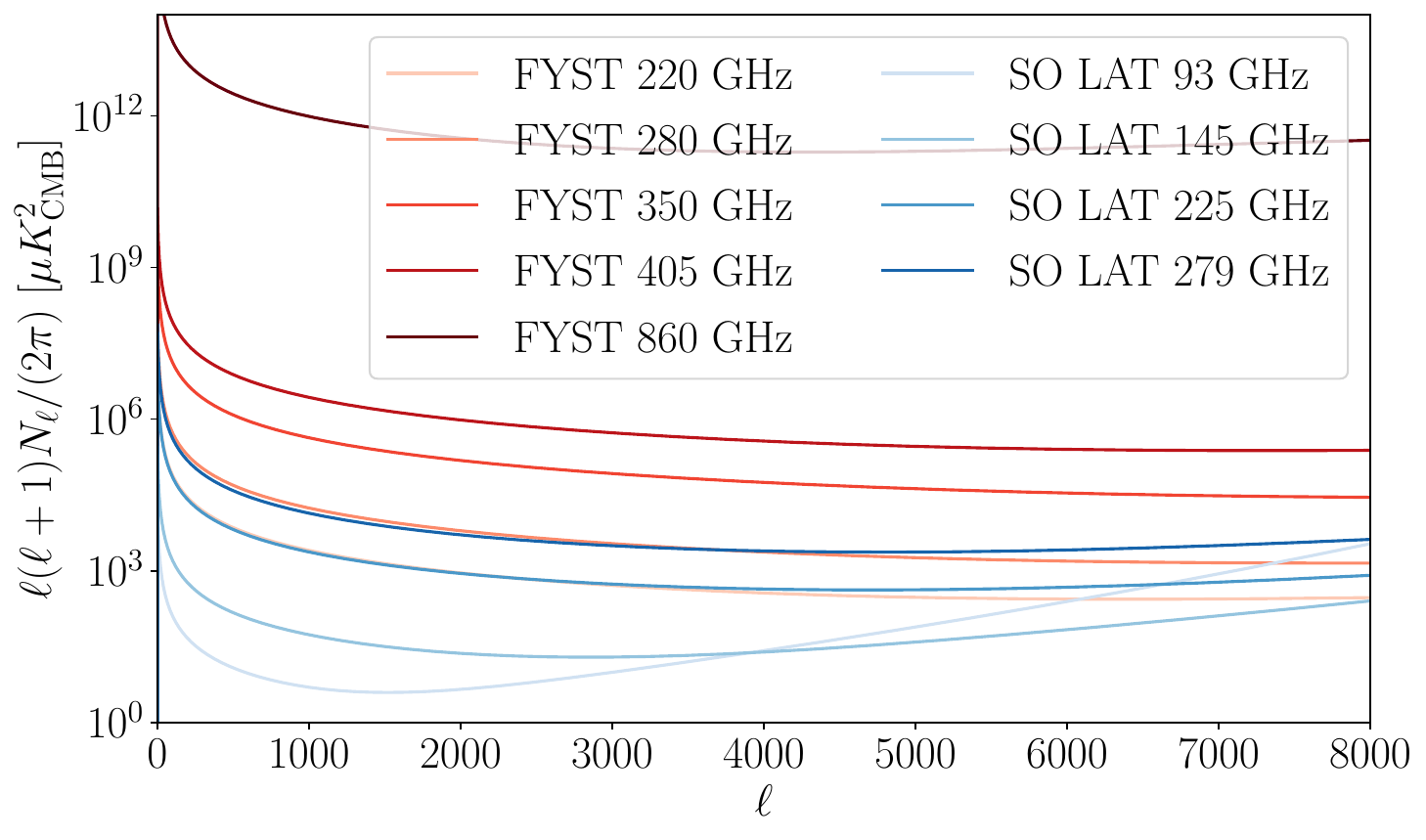}
    \caption{SO LAT (blue lines) and FYST (red lines) sum of the atmospheric red and instrumental white noise spectra based on the noise model presented by \citep{SO_2019} and adopted by \citep{Choi_2020_Noise}. The noise curves have been beam corrected.}
    \label{fig:red_noise}
\end{figure}

Our Skymodel pipeline includes two types of noise: white noise, to reproduce the instrumental noise, and atmospheric red noise, which is present for ground-based experiments. The noise implementation is based on the SO Collaboration \citep{SO_2019} noise curves forecasted and adapted for CCAT-prime by \citep{Choi_2020_Noise}:

\begin{ceqn}
\begin{align}
    N_\ell = N_\mathrm{red} \left( \frac{\ell}{\ell_\mathrm{knee}} \right)^{\alpha_{\mathrm{knee}}} + N_\mathrm{white},
\label{eqn:red_noise}
\end{align}
\end{ceqn}

where $\ell_\mathrm{knee} = 1000$ and $\alpha_\mathrm{knee} = -3.5$. The values for $N_\mathrm{red}$ and $N_\mathrm{white}$ for the SO LAT and FYST are included in the Skymodel. For this reason, the atmospheric noise model can only be computed at valid SO LAT frequencies or FYST frequencies (see Table \ref{tab:fsb}) and takes the form of a series of power spectra, as defined by Eq. (\ref{eqn:red_noise}) and shown in Fig. \ref{fig:red_noise}. Using those power spectra, the Skymodel generates full-sky map realisations of the atmospheric noise or atmospheric and instrumental noise. This is done through the HEALPy function `synfast()', which generates map realisations from the power spectrum while assuming the field to be Gaussian. This was done for each frequency band. In reality, neighbouring frequency bands will observe similar regions of the sky at the same time. Thus, the atmospheric noise is partially correlated. We adopted a $90$\% correlation between each pair of neighbouring frequency channels as suggested in \cite{SO_2019}. Fixing the `synfast()' seed allowed for the production of a $100$\% correlated map for two neighbouring channels. Another seed and a correction factor can be applied to the power spectrum to produce a down-weighted correlated map and an uncorrelated map of one of the two channels. This new map and the fully correlated map can be linearly combined to produce a $90$\% correlated map using

\begin{ceqn}
\begin{align}  
    M^{90\%}(\nu) = h(0.9^{2}C_{\ell}^{n,100\%}(\nu))+h((1-0.9)^{2}C_{\ell}^{n,0\%}(\nu)),
\label{eqn:corr_noise}
\end{align}
\end{ceqn}

where $M$ is the final map at the frequency $\nu$, which will be $90$\% correlated with its neighbouring channel. The HEALPy synfast() function is represented by $h$, and $C_{\ell}^{n}$ is the atmospheric noise curve, which is $100$\% correlated when produced from the same seed as its neighbouring channel and  $0$\% when produced from another seed. 

The frequency bands are broad and extended over some frequency range, and their transmission is not perfect but varies over their extent. In the first approximation, we considered the frequency bands to be delta functions with perfect transmission.

\begin{figure}[h!]
    \centering
    \includegraphics[angle=90,width=0.48\textwidth]{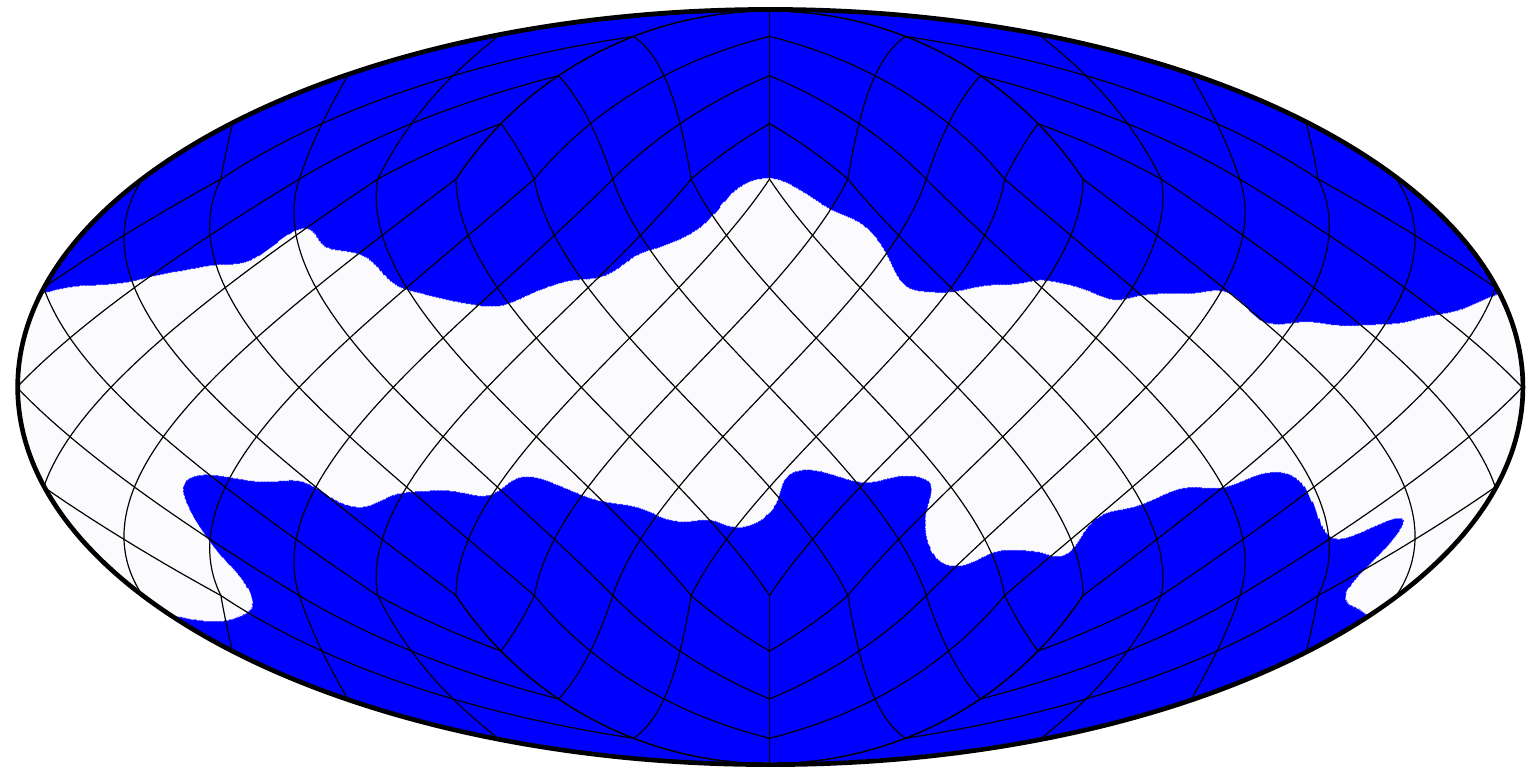}
    \caption{Galactic foreground mask derived from \textit{Planck} data by \citep{Jens_2018} ($f_\mathrm{sky} \approx 0.6$). The white part is the masked region, and the blue part is what was used in our analysis. Overplotted is the HEALPix tessellation scheme for $N_{\mathrm{side}}=4$.}
    \label{fig:masks}
\end{figure}

In addition to the maps of astrophysical emission and noise, the Skymodel uses survey masks. The masks are stored at $N_\mathrm{side} = 256$ and were upgraded to $N_\mathrm{side}=4096$ in order to correspond to the Websky map resolution. In this study, we used a mask that excludes the Galactic plane with a sky fraction of $f_{\mathrm{sky}}=0.6$ (see Fig. \ref{fig:masks}). This mask was derived in \cite{Jens_2018} by masking the 40\% brightest pixels of a \textit{Planck} Galactic emission map. This is an initial approximation, as in reality SO LAT and FYST will observe a smaller fraction of the sky.

The Skymodel pipeline also simulates the effect of telescope beams using the Gaussian FWHMs given in Table \ref{tab:fsb}. The product of the Skymodel is a full-sky HEALPix map containing all the Galactic foreground components, the extragalactic components, instrumental white noise, and atmospheric red noise at a given frequency and assuming a given beam size (See \ref{AnnexeSky}).

\section{Method to extract thermal Sunyaev-Zeldovich map}

To study the power spectrum measurement of the tSZ effect for the SO LAT and FYST telescopes, a tSZ map needs to be extracted from the simulated microwave sky. The sky is composed of various Galactic and extragalactic emissions that are considered to be contaminants of our signal of interest. Component separation methods permit the recovery of a signal from the multi-component microwave sky. 
Internal linear combination (ILC) is a  multi-frequency component separation technique that was first used by \cite{Bennett_2003} to extract CMB maps from \textit{WMAP} data and later refined by \cite{Eriksen_2004}. 
Multiple flavours of ILC exist. In their paper, \cite{Tegmark_2003} suggested applying it in Fourier space rather than in map space. Needlet ILC \citep[NILC;][]{Delabrouille_2009} processes the maps in both spatial and harmonic space by proposing to filter each multi-frequency map by a set of $\ell$-dependent window functions called needlets in order to apply multiple ILCs on different multipole intervals. This makes the ILC weights $\ell$-dependent, allowing them to take advantage of the scale-dependent distribution of astrophysical components to minimise the noise and better extract the signal of interest. The Modified ILC Algorithm \citep[MILCA;][]{Hurier_2013} differentiates itself by allowing multiple constraints on the Astrophysical emissions in order to reduce their contamination. Scale-discretised directional wavelet ILC \citep[SILC;][]{Rogers_2016} filters the map with $\ell$ windows that are direction dependent. 

In this work, we used ILC and CILC methods in a map-based space. It is also possible to use these methods directly in Fourier space, which is computationally faster but does not allow for spatial separation of signals. Fourier space ILC/CILC only has information on the power spectra of the component in $\ell$-space and no spatial information on the localisation of the signal over the full-sky map.

\subsection{Internal linear combination}

Internal linear combination is widely used in tSZ studies \citep{Bennett_2003,Fornazier_2022}. Its main advantage is that it is a so-called "blind" method because only the SED of the component of interest must be known, and no information on the contaminants is needed.
The ILC supposes that (1) the maps are composed of a linear combination of astrophysical components and noise, and
(2) the astrophysical components are uncorrelated. 
%\end{enumerate}

A set of $N_{\mathrm{obs}}$ observed map $m_{i}(p)$ at a frequency $i$ and a pixel $p$ can be expressed as

\begin{ceqn}
\begin{align}
m_{i}(p) = a_{i}s(p) + n_{i}(p),
\label{eqn:ILC_ini}
\end{align}
\end{ceqn}

where $a_{i}$ is the amplitude of the frequency spectrum of the signal of interest $s$ at each frequency $i$. The noise at a given frequency $i$ and pixel $p$ is represented as $n_{i}(p)$. 

%\begin{ceqn}
%\begin{align}
%M(p) = As(p) + N(p),
%\label{eqn:ILC_ini_vec}
%\end{align}
%\end{ceqn}

%where the vector containing all the maps at each frequency $i$, $M=[m_{i_{0}}(p),...,m_{i_{N_{\mathrm{obs}}}}(p)]$, is the linear combination of $A=[a_{i_{0}},...,a_{i_{N_{\mathrm{obs}}}}]$ which is known as the mixing vector and for which $a_{i}$ contain the amplitude of the frequency spectrum of the signal of interest $s$ at each frequency $i$. $N=[n_{i_{0}}(p),...,n_{i_{N_{\mathrm{obs}}}}(p)]$ is the vector containing all the other astrophysical components that are not of interest and the instrumental noise at each frequency $i$. 

To retrieve the signal of interest $s$, an estimator $\hat{s}_{\mathrm{ILC}}$ is defined as

\begin{ceqn}
\begin{align}
\hat{s}_{\mathrm{ILC}}(p)=\sum_{i}\omega_{i}m_{i}(p),
\label{eqn:ILC_est}
\end{align}
\end{ceqn}

where $\omega_{i}$ is the ILC weight at the frequency $i$. The ILC weights are defined such that they follow two conditions. One of them is that in order to obtain an unbiased estimate, the ILC weights must have a unit response to the frequency spectrum of the signal of interest $s$ characterised by $a_{i}$ at each frequency $i$ (see Eq. \ref{eqn:ILC_cond1}).
 
\begin{ceqn}
\begin{align}
\sum_{i}\omega_{i}a_{i}=1. 
\label{eqn:ILC_cond1}
\end{align}
\end{ceqn}

The other condition is that the minimisation of the presence of the unwanted astrophysical contaminants and the instrumental noise in the reconstructed map of the signal $s$ is achieved through the condition of minimum variance of the reconstructed map, VAR($\hat{s}_{\mathrm{ILC}}(p)$). As shown in \cite{Eriksen_2004}, because the components are uncorrelated, we have VAR($\hat{s}_{\mathrm{ILC}}(p)$)$=\sum_{k}\sum_{j}\hat{c_{jk}}\omega_{j}\omega_{k}$, where $\hat{c_{jk}}$ is the value of the covariance of the observed maps between the frequencies $j$ and $k$. Thus, the minimum variance condition is given by 

\begin{ceqn}
\begin{align}
\frac{\partial}{\partial \omega_{i}} \sum_{k}\sum_{j}\hat{c_{jk}}\omega_{j}\omega_{k}=0.
\label{eqn:ILC_cond2}
\end{align}
\end{ceqn}
    
The minimum variance under the constraint of Eq. \eqref{eqn:ILC_cond2} can be achieved using the Lagrange multiplier method to solve the above equation. The solution gives the following ILC weights: 

\begin{ceqn}
\begin{align}
\Vec{W}^{T} = \frac{\Vec{A}^{T}\Vec{\hat{C}}^{-1}}{\Vec{A}^{T}\Vec{\hat{C}}^{-1}\Vec{A}},
\label{eqn:ILC_w}
\end{align}
\end{ceqn}
where $\Vec{W}$ is the vector containing the ILC weights, and $\Vec{A}$ is the vector containing the frequency spectrum information of the signal of interest and is called `the mixing vector'. The covariance matrix of the set of maps is represented by $\Vec{\hat{C}}$. The map containing our signal of interest is given by

\begin{ceqn}
\begin{align}
\hat{s}_{ILC}(p)=\sum _{i} \omega _{i}^{T}m_{i}(p).
\label{eqn:ILC_f}
\end{align}
\end{ceqn}
From that, we can define the residual noise ($r_{N}$), that is, the noise left over by the ILC in the reconstructed map. This noise can be defined as the difference between the ILC-recovered map and the true signal : 

\begin{ceqn}
\begin{align}
r_{N} = \hat{s}_{ILC}(p) - \sum _{i} \omega _{i}^{T} a_{i}s(p) = \sum _{i} \omega _{i}^{T} n_{i}(p).
\label{eqn:ILC_residual}
\end{align}
\end{ceqn}

The contaminants are not absent from the recovered signal map. Moreover, the low variance contaminants might not be reduced in the resulting map. This becomes a problem when assumption two of the ILC is not fully verified and our target signal is spatially correlated with another emission, such as for the tSZ and CIB. This is the case between CIB and tSZ. The remaining CIB signal in the map can bias the ILC estimate of the tSZ. One way to reduce this effect is through CILC.

\subsection{Constrained internal linear combination}

The CILC method was first introduced by \cite{Remazeilles_2011}. It adds one extra constraint to the ILC approach. If the spectral shape of one of the contaminants is known, it can be taken into account in the modelling such that instead of Eq. \eqref{eqn:ILC_ini}, it is supposed that  

\begin{ceqn}
\begin{align}
m_{i}(p) = a_{i}s(p) + b_{i}q(p) + n_{i}(p),
\label{eqn:CILC_ini_vec}
\end{align}
\end{ceqn}
where $s(p)$ is the signal of interest and $q(p)$ is the contaminant we want to suppress. Respectively, $a_{i}$ and $b_{i}$ are the intensity of the frequency spectrum evaluated at a frequency $i$ of the signal and contaminant. All the remaining astrophysical contaminants and the instrumental noise at a given frequency $i$ and pixel $p$ are contained by  $n_{i}(p)$. The CILC weights are defined such that 

\begin{ceqn}
\begin{align}
\sum_{i}\omega_{i}a_{i} = 1 \ \mathrm{and} \ \sum_{i}\omega_{i}b_{i}= 0.
\label{eqn:CILC_cond1}
\end{align}
\end{ceqn}

The CILC weights have a unit response to the SED of the signal of interest $s$ and a null response to the SED of the astrophysical contaminant $q$ we want to suppress so that this contaminant is absent from the reconstructed signal. 

The condition of minimum variance (see Eq. \eqref{eqn:ILC_cond2}) stays the same, and only the expression of the weights is changed to take into account the two conditions imposed by the CILC,

\begin{ceqn}
\begin{align}
\Vec{W}^{T} = \frac{(\Vec{B}^{T}\Vec{\hat{C}}^{-1}\Vec{B})\Vec{A}^{T}\Vec{\hat{C}}^{-1}-(\Vec{A}^{T}\Vec{\hat{C}}^{-1}\Vec{B})\Vec{B}^{T}\Vec{\hat{C}}^{-1}}{(\Vec{A}^{T}\Vec{\hat{C}}^{-1}\Vec{A})(\Vec{B}^{T}\Vec{\hat{C}}^{-1}\Vec{B})-(\Vec{A}^{T}\Vec{\hat{C}}^{-1}\Vec{B})^{2}},
\label{eqn:CILC_w}
\end{align}
\end{ceqn}

where $\Vec{W}$ is the vector containing the CILC weights, $\Vec{B}$ is the mixing vector of the contaminant, $\Vec{A}$ is the mixing vector of the signal, and $\Vec{\hat{C}}$ is the covariance matrix of the set of maps. Those weights are then applied to the set of multi-frequency maps using Eq. \eqref{eqn:ILC_f} to recover a map containing the signal of interest free from the nulled contaminant and with minimum contamination from other astrophysical signals and instrumental noise. The CILC method can be generalised to null more than one contaminant if their SED is known. However, these additional constraints reduce the degree of freedom available for the condition of minimum variance to be verified \citep{Hurier_2013}, thus generating increased noise for the CILC compared to the ILC. Similar to the ILC, the residual noise can be defined as the difference between the CILC-recovered map and the signal of interest (see Eq. \ref{eqn:ILC_residual}), except that the weights are given by Eq. \eqref{eqn:CILC_w}.

\subsection{Beam convolution}

The set of multi-frequency maps was smoothed with a Gaussian kernel to the lowest frequency channel resolution so that they could be linearly combined by the ILC. No meaningful physical information is available on scales smaller than the resolution of each map. When linearly combining the maps, the upper limit of meaningful physical information was therefore given by the lowest resolution map.  

More sophisticated component separation methods, such as NILC, allows the original resolution of each frequency map to be kept. Indeed, the NILC weights are a function of the multipole $\ell$ for each frequency map and can therefore be set to zero when the scale exceeds the resolution of the map. Setting the weights to 0 above a given $\ell$, allows the lower-resolution maps to contribute to the signal reconstruction only up to their beam-limited scale, leaving the higher-resolution maps to be the contributors of signal reconstruction on smaller scales.

We note that when simulating the microwave sky, only the astrophysical components and atmospheric noise are limited by the instrument beam. Instrumental noise is not beam limited. The inclusion of the beam was done using HEALPy to smooth the maps $a_{\ell,m}$ in the multipole space and multiply them with a $\ell$-dependent Gaussian kernel. Our set of multi-frequency maps can be expressed as

\begin{ceqn}
\begin{align}
a_{\ell,m}(m_{i}^{s}) = a_{i}{b_{i}(\ell)}a_{\ell,m}(s) + {b_{i}(\ell)}a_{\ell,m}(c_{i}) + a_{\ell,m}(n_{i}),
\label{eqn:ILC+beam}
\end{align}
\end{ceqn}
where $a_{\ell,m}(m_{i}^{s})$ are the $a_{\ell,m}$ coefficient of the smoothed map at the frequency $i$. The Gaussian beam in the multipole space at the frequency $i$ is indicated as $b_{i}(l).$  Next, $a_{\ell,m}(s)$ is the $a_{\ell,m}$ decomposition of component of interest map, and $a_{\ell,m}(c_{i})$ represents all astrophysical contaminants and the atmospheric noise at a frequency $i$. Finally, $a_{\ell,m}(n_{i})$ is the instrumental noise $a_{\ell,m}$ at the frequency $i$. In general, to apply an ILC, all maps have to be smoothed down to the lowest resolution. This is necessary to avoid adding information from other higher resolution, higher frequencies maps that are smaller than the beam size of some of the maps into the recovered map. The smoothing of the maps to the lowest map resolution was done using HEALPy, and this process can be formalised as

\begin{ceqn}
\begin{align}
a_{\ell,m}(m_{i}^{*}) = a_{i}b_{\mathrm{f}}(\ell)a_{\ell,m}(s) + b_{\mathrm{f}}(\ell)a_{\ell,m}(c) +\frac{b_{\mathrm{f}}(\ell)}{b_{i}(\ell)}a_{\ell,m}(n_{i}),
\label{eqn:ILC+beam2}
\end{align}
\end{ceqn}
where $a_{\ell,m}(m_{i}^{*})=(b^{\mathrm{f}}(\ell)/b_{i}(\ell))a_{\ell,m}(m_{i}^{s})$ are the $a_{\ell,m}$ coefficient of the map at a frequency $i$ smoothed down to the lowest resolution. The Gaussian kernel at this desired `lowest' resolution is indicated at  $b^{\mathrm{f}}$. The consequence of the necessary smoothing to the lowest resolution is an increase of the instrumental white noise contribution in each frequency map by a factor $b_{\mathrm{f}}(\ell)/b_{i}(\ell)$ because for all the frequency channels, except the lowest one, we have $b_{\mathrm{f}}(\ell) \textgreater b_{i}(\ell) $.  

\subsection{Full-sky treatment}

The microwave full-sky maps are composed of many different astrophysical signals. Some of those signals are spatially localised. The Galactic foregrounds are mainly localised around the Galactic plane. When applying the map-based ILC to a full-sky map, instead of applying it straightforwardly to the multi-frequency set of maps, thus getting only one set of weights for the full sky and equally treating regions that are strongly contaminated by Galactic foregrounds and regions that are less contaminated, we tessellated the sky into different zones and applied an ILC on each of these zones separately. This resulted in a set of different weights for each region so that they are always optimised for the level of contamination present in their zone. This has been done previously in various studies \citep{Tegmark_2003}. The most common tessellation approach is to separate the sky into zones of Galactic contamination. 

To tessellate the sky, we used the native nested HEALPix tessellation (see Fig. \ref{fig:masks}). This nested scheme tessellates the sky originally into 12 equal-area fields. The sides of each of those fields can be further divided to create more equal-area cells of smaller sizes. Using HEALPix nested scheme to create sub-fields \citep{Healpy_2005}, we chose to tessellate the sky using $N_\mathrm{side} = 4$, which gives $N_\mathrm{pix} = 12\times N_\mathrm{side}^{2}=192$ fields, each covering an area of $\approx 215 \mathrm{deg}^{2}$. In general, tessellation can create border effects along the limits of the fields. In \cite{Eriksen_2004}, the border effects were attenuated by smoothing the ILC weights. In this work, we do not correct the border effects. 

The Galactic plane was masked (see Fig. \ref{fig:masks}) with a  $f_ {sky} \approx 0.6$ Galactic mask map to reduce foreground contamination and better extract the tSZ effect. To compute power spectra on a masked map, we used PyMaster \citep{Alonso_2019}, which corrects the power spectra for the masking of part of the sky. In this work, we do not apodise the masks, as it is computationally very intensive at our high resolution and does not make a significant difference in the resulting power spectrum (see Fig. \ref{fig:tess_mask_apo}).
The full treatment which includes, creation, tessellation, masking and ILC is performed at the map level. Only when a tSZ full-sky map was extracted was the power spectrum of the map computed using spherical harmonics.  

\subsection{Debiasing} \label{sec_debias}

The instrumental white noise biasing of the tSZ power spectrum extraction was estimated for a $4000$ h observation time \citep{Choi_2020}. Therefore, the mean of this noise could be computed and removed to debias the tSZ reconstruction. This was done by generating random realisation of the instrumental white noise for each frequency at the sensitivities given in Table \ref{tab:fsb} and computing the projected noise by multiplying the ILC weights from the tSZ reconstruction with the noise maps and summing them. The power spectrum of this noise was then subtracted from the ILC-extracted Compton-y ($y_{\mathrm{ILC}}$) power spectrum. All the following power spectrum results were debiased for instrumental white noise. 

Similarly, as the CMB power spectrum is well-known, we could generate random realisation from the Websky CMB power spectrum to debias for the CMB residual noise. In the first approximation, this study presents an idealised case where the CMB power spectrum is perfectly known and its residual is removed. With real data, this CMB debiasing will be limited by the uncertainties on the recovered CMB power spectrum. 

\subsection{Overview of steps}

\begin{figure*}[h!]
    \centering
    \includegraphics[width=0.8\textwidth]{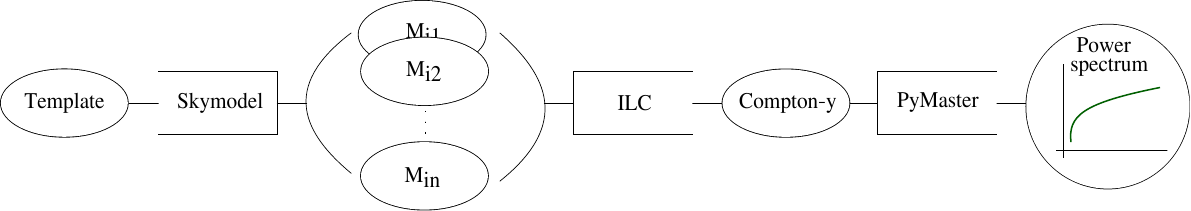}
    \caption{Scheme of the method we use in this paper to extract the tSZ signal. The input of the sky model is template maps coming from the Websky simulations and PySM. The sky model processes them to produce maps of the different sky components at a given frequency, sensitivity, and beam. This set of multi-frequency maps $         \{ \mathrm{M}_{\mathrm{i1}},...,\mathrm{M}_{\mathrm{in}} \}$ was smoothed to a common resolution and tessellated, and the Compton-y signal was extracted by the ILC. The PyMaster algorithm was used on the resulting masked map to extract the power spectrum.}
    \label{fig:scheme_method}
\end{figure*}

In this section, we describe our tSZ extraction method and illustrate it in Fig. \ref{fig:scheme_method}. 

First, using the Skymodel based on the Websky simulation and PySM, we generated multi-frequency maps of the microwave sky at the frequencies, sensitivities, and beam sizes of SO LAT and FYST (see Table \ref{tab:fsb}). The set of maps is in the HEALPix format and uses intensity $\mathrm{MJy \, sr}^{-1}$ units. 

Second, the simulated maps were all smoothed down to the resolution of the lowest frequency channel used in our analysis.

Third, a Galactic mask that excludes 40\% of the sky (see Fig. \ref{fig:masks}) was applied to the set of multi-frequency maps.

Fourth, each map was tessellated into smaller equal-area regions using the HEALPix nested format. 

Fith, an ILC was applied to each multi-frequency set of the tessellated area to obtain the Compton-y patches. The patches were reassembled to obtain the full-sky Compton-y map. The covariance matrix was computed only on the unmasked data by our ILC.

Sixth, the power spectrum of the recovered Compton-y map was computed using PyMaster and debiasing, and the residual noises were computed using Eq.\eqref{eqn:ILC_residual}

\section{Results}

To quantify how much CIB could leak into the reconstructed tSZ maps for future experiments and access the detectability of the tSZ power spectrum, we used the microwave sky model detailed in Sec. \ref{des_skymodel}. We simulated a set of multi-frequency maps at the frequencies, sensitivities, and beam sizes for SO LAT and FYST (see Table \ref{tab:fsb}) with atmospheric noise. We applied an ILC or CILC component separation method to extract a tSZ map and looked at the ratios between the different noises and signal power spectra. We restricted ourselves to the scales $\ell \in [500,5000]$. We chose this lower limit to avoid the cosmic variance and the upper limit because of the $2.2'$ resolution of our HEALPix maps, which allowed us to resolve scales up to $\ell \sim 4909$.

\subsection{Internal linear combination results} \label{ILC_results}

\begin{figure}[h!]
    \centering
    \includegraphics[width=0.45\textwidth]{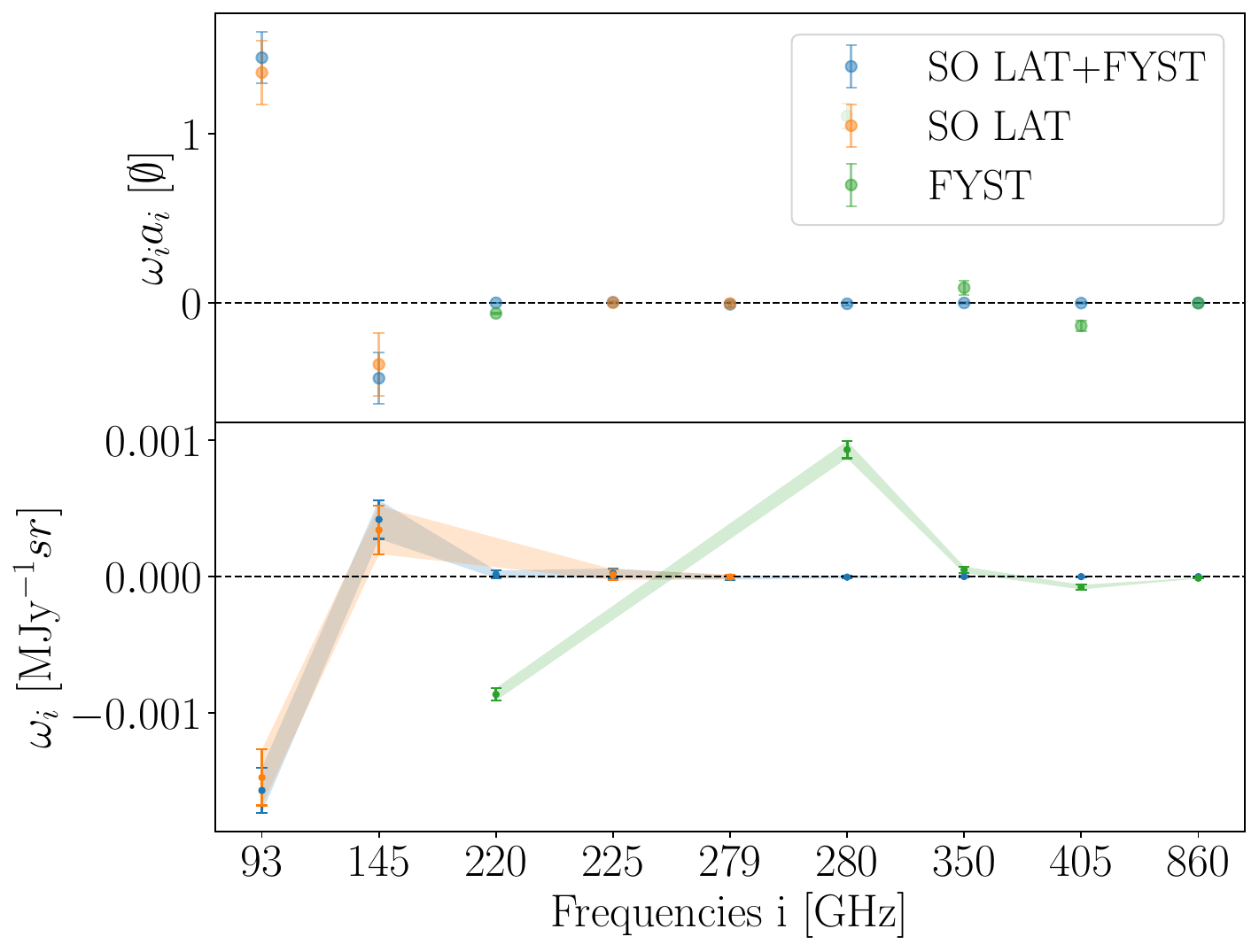}
    \caption{Internal linear combination weights for SO LAT-like (orange) and FYST-like (green) experiments and SO LAT+FYST (blue) when the sky contains all extragalactic components (see Table \ref{tab:skymodel}), Galactic foregrounds, and atmospheric noise. Top figure:  Fractional contribution of each frequency channel to the recovered $y$-map. Bottom figure: Frequency channels that contribute to maximising the tSZ signal and minimising the noise. The error bars show the standard deviation around the mean weight of all of the $192$ tessellated fields for each frequency map.}
    \label{fig:w_Atmo}
\end{figure}

By construction, the weights of an ILC must verify the first condition, which is to have a unit response to the SED of the tSZ effect (see Eq. \eqref{eqn:ILC_cond1}). This first condition means that the element-wise product of weights ($\omega_{i}$) times the mixing vector ($a_{i}$) is an indicator of the fractional contribution of each frequency band to the recovery of the tSZ signal. It can be negative since the frequency spectrum of the tSZ effect is negative below $217$ GHz. The weights attributed by the ILC can themselves be negative for optimisation purposes, with the only constraint being that the sum over all frequencies must be one. The value of the weights is given by the mean of the 192 weights (one per tessellated field), and the error bars are given by the standard deviation of those weights. The top panel of Fig. \ref{fig:w_Atmo} shows that for SO LAT alone (orange dots), the 93 GHz channel is the main contributor to the tSZ reconstruction by a factor $\omega_{93}a_{93}\approx 1.4$. This channel overestimates the tSZ signal, and the SO LAT 145 GHz channel was used by the ILC to correct this. As a result, the 145 GHz channel contributes roughly by a factor $\omega_{145}a_{145}\approx -0.4$ to the tSZ reconstruction. This result is not surprising, as those frequency bands are located at the tSZ decrement that is characteristic of the signal in the microwave sky and where the values of the mixing vector are negative. When SO LAT and FYST are combined (SO LAT+FYST; blue dots), the SO LAT 93 GHz and 145 GHz channels remain the main contributors to the tSZ reconstruction and have similar values, $\omega_{93}a_{93}\approx 1.45$ and $\omega_{145}a_{145}\approx -0.45$. Other channels from SO LAT and FYST do not contribute to the reconstruction. When FYST is alone (green dots), the main contributor to the tSZ reconstruction is the 280 GHz channel, by a factor $\omega_{280}a_{280}\approx 1.1$, and the 350 GHz channel, by $\omega_{350}a_{350}\approx 0.05$. The 220 GHz and 405 GHz channels were used to mitigate the over evaluation of the tSZ signal with the respective factors $\omega_{220}a_{220}\approx -0.05$ and $\omega_{405}a_{405}\approx -0.1$. 

The second condition that has to be verified by the ILC weights is to ensure that the variance of the recovered map has been minimised (see Eq. \eqref{eqn:ILC_cond2}). The bottom panel of Fig. \ref{fig:w_Atmo} shows the values of the weights. In combination with the top panel, we concluded that the weights associated with some frequency bands that do not contribute to the tSZ signal reconstruction but have a high absolute value contribute to the noise minimisation. In particular, the 220 GHz channel located close to the tSZ null is often attributed a high weight by the ILC even though it does not contribute to the tSZ reconstruction. This is because it is the only map containing all the noise contaminants and almost no signal. It is thus the perfect map for noise minimisation. But this is not the case in Fig. \ref{fig:w_Atmo}, except for when FYST is alone (green dots), because of the presence of atmospheric noise in our mock sky. The strong atmospheric noise contaminates the 220 GHz band, making it less exploitable by the ILC (see the comparison with Fig. \ref{fig:w_Planck} for a case without atmospheric noise). Figure \ref{fig:w_Atmo} shows that in the presence of all the microwave emissions and atmospheric noise, the SO LAT frequency channels that probe the tSZ decrement, 93 GHz and 145 GHz, drive both the tSZ reconstruction and the noise minimisation of the tSZ signal.

\begin{figure*}[h!]
    \centering
    \includegraphics[width=1\textwidth]{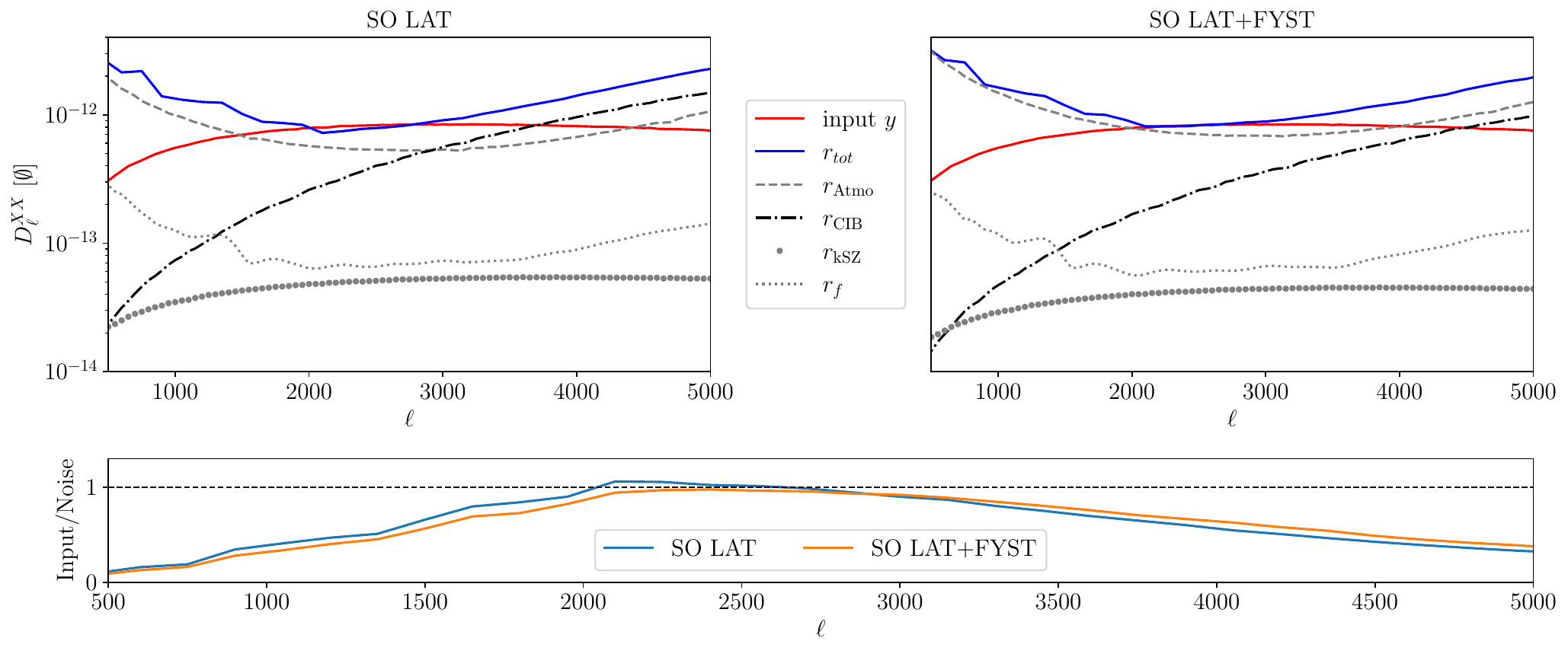}
    \caption{Power spectra of the ILC residual noises compared to the input Compton-y power spectrum from the Websky simulations (red). The simulated sky contains all extragalactic components (see Table \ref{tab:skymodel}), Galactic foregrounds, and atmospheric noise. Top panels: Power spectra of various quantities. The grey curves show the ILC noises residual power spectra, denoted by $r_{\mathrm{X}}$ (X being the kSZ); the cumulative Galactic foregrounds ($f$); and the atmospheric noise (Atmo). The black curve shows the power spectrum of the CIB residual noise. The blue curve shows the sum of all grey and black curves, that is, the total ILC noise residual debiased for instrumental white noise and the CMB. 
    The top-left panel shows the results for the simulated SO LAT data, while the top-right panel shows the results for the simulated data for SO LAT and FYST combined. Bottom panel: Ratio of the input Compton-y power spectrum (red) and the total noise residual (blue) for SO LAT and SO LAT+FYST. All power spectra have been beam corrected and bin averaged over a window $\Delta \ell=50$.}
    \label{fig:PS_Atmo_debias}
\end{figure*}

However, the weights do not tell us anything about the CIB contamination or the detectability of the tSZ signal. For that, we turned to the power spectrum and the residual ILC noise power spectra. The ILC output is a map containing the Compton-y signal and residual noises (Eq. \ref{eqn:ILC_residual}). Using PyMaster, the power spectra of this map and the residual noises can be computed and compared to the input-pure Compton-y map power spectra from the Websky simulations. We defined $D_{\ell}^{XX} = \ell(\ell+1)C_{\ell}^{XX}/2\pi$ as the power spectra of the quantity $X$  or of the residual noise $r_{X}$ of the quantity $X$. For SO LAT alone (see Fig. \ref{fig:PS_Atmo_debias} top-left panel) the dominant noise residual on small scales is the CIB residual ($r_{\mathrm{CIB}}$), which starts to dominate around $\ell \sim 3700$. The cumulative foreground residual ($r_{f}$) and the kSZ residual ($r_{\mathrm{kSZ}}$) are both subdominants to the input signal by around one order of magnitude over the scales $\ell \in [500,5000]$.
The atmospheric noise residual ($r_{\mathrm{Atmo}}$) dominates the input signal for $\ell < 1400$ and $\ell > 4500$. 

In the combined case of SO LAT+FYST (see right panel of Fig. \ref{fig:PS_Atmo_debias}), the situation is similar to the case where SO LAT is alone except that the atmospheric noise residual ($r_{\mathrm{Atmo}}$) is higher when FYST is present. This could be due to higher atmospheric noise being present in the high-frequency channels of FYST (see Fig. \ref{fig:red_noise}). The CIB noise residual power spectrum is significantly lower for SO LAT+FYST. It starts to dominate above the input signal around $\ell \sim 4200,$ while for SO LAT alone it was around  $\ell \sim 3700$. We concluded that adding FYST to SO LAT reduces the CIB noise residual.

The lower panel of Fig. \ref{fig:PS_Atmo_debias} shows the extent the input Compton-y power spectrum is dominated by the total ILC noise residual as a function of the scale. 
The total noise residual ($r_{tot}$) power spectrum dominates over the input signal on nearly all scales except around $\ell \in [2100,2700]$
for SO LAT alone. This is because of our debiasing of the CMB on those critical scales. Overall, SO LAT alone has a better input/noise ratio on scales $\ell<2500$ because of its slightly lower atmospheric noise residual, and SO LAT+FYST does better than SO LAT alone on scales $\ell>3000$ because of its lower CIB noise residual.

\subsection{Constrained internal linear combination results} \label{CILC_results}

Constrained internal linear combination can be used to ‘null’ the contribution of a component of the microwave sky using its frequency spectrum. Nullifying several components was tested. 

Before any debiasing, the CMB is the main contaminant of the recovered Compton-y power spectra on large scales, so it seemed natural to deproject (i.e. ‘null’) it. However, the additional constraints on the ILC weights (see Eq. \eqref{eqn:CILC_ini_vec}) led to a smaller solution space for the weights to extract the signal, which resulted in a ‘noise penalty’, that is, higher noise in the recovered signal map. In this particular case of a mock sky containing all Galactic foregrounds and extragalactic emission, excluding RPSs, and applying a CILC to deproject CMB, the ILC weights are so constrained to null CMB that they become less efficient at reducing the second main contaminant, the atmospheric noise contamination, which blows up. As a consequence, the total residual noise dominates the input signal even more, by up to one order of magnitude on small scales (see Annexe \ref{AnnexeCILC}). Deprojecting the CMB is therefore not a recommended option for our study. This effect was already reported in Fig. 36 of \cite{SO_2019}, but in their study, the blowing up of the noise was mitigated on large scales by the addition of \textit{Planck} data.

Because of its high-frequency coverage, we expected FYST to better probe the CIB, which is the dominant contaminant above $\nu > 300$ GHz, and therefore better target CIB contamination and remove it. Deprojecting the CIB using CILC seemed the natural option to take full advantage of the potential constraining power of the FYST high-frequency channels. We note that a perfect deprojection of the CIB is not possible because its signal is not composed of only one SED. We adopted the modified blackbody spectrum given in Eq. \eqref{eqn:mbb_sed}, with $\mathrm{A}_{\mathrm{CIB}}=1$, $T_{\mathrm{dust}}=10$ K, and $\beta =1.6$. Parameter values were calculated by computing the average for each of the parameter maps derived in Sec.\ref{sec_extragal}.  

\begin{figure*}[h!]
    \centering
    \includegraphics[width=1\textwidth]{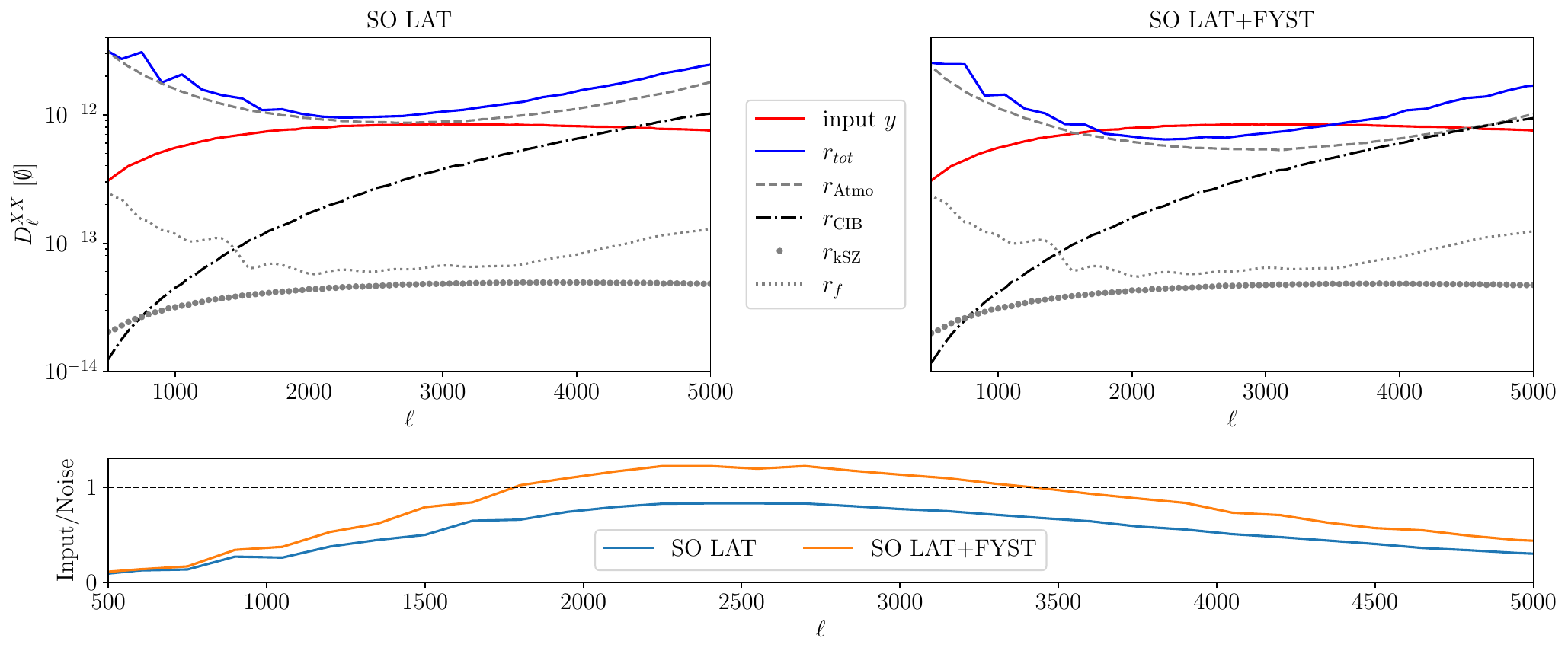}
    \caption{Power spectra of the CILC residual noises when deprojecting CIB compared to the input Compton-y power spectrum from the Websky simulations (red). The simulated sky contains all extragalactic components (see Table \ref{tab:skymodel}), Galactic foregrounds, and atmospheric noise. Top panels: Power spectra of various quantities. The grey curves show the ILC noise residual power spectra, denoted by $r_{\mathrm{X}}$ (X being the kSZ); the Galactic foregrounds ($f$); and the atmospheric noise (Atmo). The black curve shows the power spectrum of the CIB residual noise. The blue curve shows the sum of all grey and black curves, that is, the total ILC noise residual debiased for instrumental white noise and the CMB. The top-left panel shows the results for the simulated SO LAT data, while the top-right panel shows the results for the simulated data for SO LAT and FYST combined. Bottom panel: Ratio between the input Compton-y power spectrum (red) and the total noise residual (blue) for SO LAT and SO LAT+FYST. All power spectra have been beam corrected and bin averaged over a window $\Delta \ell=50$.}
    \label{fig:PS_Atmo_CILC}
\end{figure*}

The top panel of Figure \ref{fig:PS_Atmo_CILC} shows that the CIB noise residual ($r_{\mathrm{CIB}}$) power spectrum is reduced by the CILC deprojection for both SO LAT and SO LAT+FYST, but it is not zero because the CIB is composed of multiples spectra and only one is used for the deprojection. It also shows that the atmospheric noise residual ($r_{\mathrm{Atmo}}$) power spectrum increases for SO LAT and decreases for the combined SO LAT+FYST when the CIB is deprojected (see Fig. \ref{fig:PS_Atmo_debias} for comparison). This can be explained by the higher degrees of freedom available with the higher number of frequency channels for SO LAT+FYST, making the CILC noise penalty have a less penalising effect in comparison to when SO LAT is alone. The lower panel of the figure shows that SO LAT+FYST offers a large $\ell \in [1700,3500]$ window over which the Compton-y signal is not noise dominated, while for SO LAT alone, the signal remains noise dominated (exclusively by the atmospheric noise). For SO LAT alone or for SO LAT+FYST combined, compared to the ILC case, the ratio between the input target signal and the resulting noise is not much higher in general (Fig.\ref{fig:PS_Atmo_debias}). This was also seen when applying our pipeline to \textit{Planck} data, as the recovered Compton-y map amplitude is roughly the same as the ILC and CILC. This is not surprising because we also saw that the ILC and CILC weights are very similar. This observation of the CILC not providing a much lower noise bias Compton-y map compared to the ILC was already reported by \cite{Alonso_2018}. 

\subsection{Overview}

\begin{figure}[h!]
    \centering
    \includegraphics[width=0.45\textwidth]{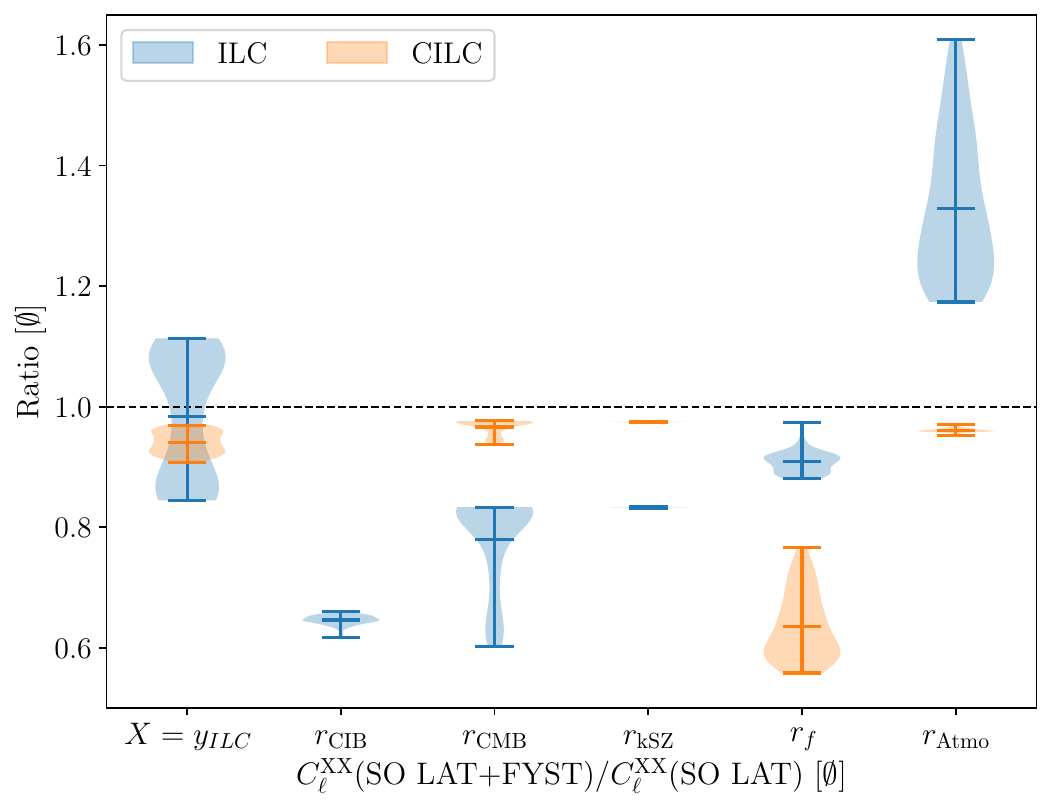}
    \caption{Value of the ratios of the power spectra of X for SO LAT+FYST compared to SO LAT($C_{\ell}^{XX}(\mathrm{SO\ LAT+FYST})/C_{\ell}^{XX}(\mathrm{SO\ LAT})$). The ILC-extracted Compton-y signal ($y_{\mathrm{ILC}}$) and $r_{\mathrm{X}}$ represent the residual noise of the quantity X. The shaded regions are the density distribution of the ratios over the $\ell$ range. The bar is the mean averaged over the $\ell$ range of the ratio distribution. In blue is the ILC case (see Sec. \ref{ILC_results}) and in orange is the CILC case when CIB is deprojected (see Sec. \ref{CILC_results})}
    \label{fig:ratio_final}
\end{figure}

Figure \ref{fig:ratio_final} gives an overview of the relative amplitude changes for the power spectrum of residual noises of each astrophysical component when combining SO LAT and FYST and shows a comparison to SO LAT alone in the two previous cases applying an ILC (in blue) and a CILC deprojecting the CIB (in orange). To see how much SO LAT+FYST reduces residual noise power spectra in comparison to SO LAT alone, we took the ratio of the residual power spectra of each component in both cases:

\begin{ceqn}
\begin{align}
R = \frac{C_{\ell}^{XX}(\mathrm{SO\ LAT+FYST})}{C_{\ell}^{XX}(\mathrm{SO\ LAT})},
\label{eqn:ILC+beam2}
\end{align}
\end{ceqn}

where $C_{\ell}^{XX}(\mathrm{SO\ LAT+FYST})$ is the power spectrum of the component $X$ or of the residual noise $r_{X}$ for SO LAT + FYST and $C_{\ell}^{XX}(\mathrm{SO\ LAT})$ is the same quantity but for SO LAT alone. For the extracted Compton-y map ($y_{\mathrm{ILC}}$), the blue violin plot shows a mean improvement of around $ 4\%$ for SO LAT+FYST compared to SO LAT alone but with a large scatter. Especially at $\ell < 2800$, SO LAT performs better (see bottom panel of Fig. \ref{fig:PS_Atmo_debias}). When deprojecting the CIB (orange violin plot), we moved to a mean improvement of around $ 8\%$ in the ILC-extracted Compton-y map for SO LAT+FYST compared to SO LAT alone. In this case, the scatter is also much smaller than before because of the high number of degrees of freedom offered by the combined SO LAT+FYST frequency bands, and  the CILC noise penalty is much lower for SO LAT+FYST compared to SO LAT alone, making the combination strictly superior to extract a less biased tSZ power spectrum in the CILC case. 

For the residual CIB noise ($r_{\mathrm{CIB}}$), the power spectrum is on average around $35\%$ lower when combining SO LAT and FYST in the ILC case. In the CILC case, it drops to $\sim 7\%$, which might not be surprising because the CILC focusses on nulling as much CIB as possible no matter the frequencies. The addition of FYST does reduce the CIB residual even more but not as much as without the null constraint. Another reason for the drop in improvement when combining SO LAT and FYST in the CILC case might be the higher atmospheric noise at the high frequencies where the CIB signal dominates, undermining the advantage of FYST to constrain and remove CIB.

In the ILC case, the CMB residual noise ($r_{\mathrm{CMB}}$) power spectrum is reduced by $\sim 22\%$ when combining SO LAT with FYST compared to SO LAT alone. Because the ILC cannot separate CMB and kSZ due to their identical spectral frequency distribution, the residual kSZ noise ($r_{\mathrm{kSZ}}$) is also suppressed when combining SO LAT and FYST, on average by around $ 18\%$. When deprojecting CIB with a CILC, adding FYST to SO LAT also reduces the CMB and kSZ residual noises but only by $\sim 5\%$ over all scales on average. 

The sum of all galactic foreground noise residuals is defined as the cumulative foreground residual ($r_{f}$). Its noise power spectrum is improved on average by $\sim 10\%$ for SO LAT+FYST compared to SO LAT in the ILC case. In the CILC case, the cumulative foreground residual power spectrum is reduced by $\sim 30\%$ when combining FYST with SO LAT. This could be due to the additional constraint on the CIB dust spectrum, nulling part of the Galactic dust contribution, which is more easily nulled with FYST high-frequency coverage. 

The FYST high-frequency channels are more contaminated by the atmosphere. For this reason, the atmospheric red noise residual ($r_{\mathrm{Atmo}}$) power spectrum is $\sim 30\%$ higher when SO LAT and FYST are combined rather than when SO LAT is alone in the ILC case. When deprojecting the CIB, the atmospheric noise residual becomes slightly lower $\sim 1\%$ for SO LAT+FYST compared to SO LAT alone. This is due to the noise penalty generated by the additional constraint in the CILC worsening the atmospheric noise for  LAT alone compared to SO LAT+FYST. Because SO LAT+FYST  benefits from additional frequency bands and therefore more degrees of freedom, it is more resilient to the inevitable noise increase caused by additional spectral constraints.

\section{Summary and discussion}

Using full-sky HEALPix template maps from the Websky simulations \citep{CITA_2020} and PySM, we created a pipeline that allowed us to simulate the high-resolution maps of the different microwave sky emissions at any beam, frequency, and sensitivity. Based on the \cite{Choi_2020_Noise} noise modelling, our pipeline also simulates atmospheric red noise at the locations of the Simons Observatory and FYST.

We simulated the sky that would be seen during experiments at these locations and used a map-based ILC and CILC to extract back the thermal Sunyaev-Zeldovich (tSZ) effect from this multi-frequency, multi-component sky. To minimise Galactic foreground contamination, we used a Galactic mask derived from \textit{Planck} (See Fig. \ref{fig:masks}). To optimise our ILC weights in order to take advantage of the spatial distribution of some of the microwave emissions, we tessellated the sky using the HEALPix equal-area nested tessellation scheme. We made use of CILC to deproject the CIB by using a simple modified blackbody spectrum to represent the CIB frequency variation and constrain the ILC weights to nullify it. We then used PyMaster to compute the power spectrum and its associated residual noise power spectra, which are the noise coming from all the other microwave sky emissions left by ILC and CILC in the recovered map.

We note that our study does have certain shortcomings that should be taken into account. For example, the Websky simulations do not yet offer a map of RPSs, and these extragalactic components are therefore absent from our study. The RPSs are an important contaminant of the tSZ signal, as they are located in galaxy clusters. If RPSs are not  considered, they can fill up the tSZ decrement and bias an estimation towards a lower estimated tSZ signal. A solution to this problem is to mask the RPSs, but this also lowers the estimated tSZ power spectrum, as it removes parts of the tSZ signal. This effect especially impacts measurements of the tSZ signal below 90 GHz, which we do not study here. A study from \cite{Holder_2002} showed that RPS subtraction can lead to an underestimation of the tSZ power spectrum of between $\sim 30\%$ to $\sim 50\%$, depending on the scale. We did not consider the relativistic correction of the tSZ effect, which can lead to a significant underestimation of the signal \citep{Remazeilles_2018}. This correction is left to future studies. Regarding the method, our tessellation scheme can lead to border effects that were reported by \cite{Eriksen_2004} but not accounted for. Our pipeline was tested on \textit{Planck} data with and without tessellation and only a marginal (i.e. $<5\%)$ difference could be seen in the power spectrum amplitude. Applying ILC in map-based space also has the disadvantage of the necessary resolution downgrade of all maps to the lower beam channel, which is 2.2' for SO LAT, thus reducing the benefits of FYST's higher resolution channels. This also prevented us from using \textit{Planck} data that are essential to lowering atmospheric and CMB contamination on large scales. This was addressed by performing CMB debiasing of the residual CMB noise in the extracted Compton-y map. A Fourier space or needlet ILC implementation is left for future work.    

We find that when using an ILC to extract tSZ, the combination of SO LAT with FYST reduces the CIB residual noise power spectrum by $\sim 35\%$,  on average, on scales $\ell \in [500,5000]$ when compared to SO LAT alone. For the CMB and kSZ components, the addition of FYST to SO LAT reduces the residual noise by  $\sim 22\%$ on average, with a strong scale dependence for the CMB, as the addition of FYST mainly helped on large scales. For the kSZ component, the reduction observed when adding FYST to SO LAT is not scale dependent. Adding FYST also helps reduce the Galactic foreground contamination by, on average, $\sim 10\%$ on scales $\ell \in [500,5000]$. However, FYST's high-frequency channels are more affected by atmospheric noise due to poorer transmission. Therefore, the residual noise power spectrum coming from the atmosphere is on average $30\%$ higher when SO LAT and FYST are combined compared to when SO LAT is alone. This greatly affects the total residual noise budget, lowering the benefits of adding FYST to SO LAT for the tSZ recovery to only an average $\sim 4\%$, and on larger scales, SO LAT alone performs better. We find that over $\ell \in [2100,2700]$, the ILC-extracted tSZ power spectrum is not noise dominated for SO LAT alone. But on smaller scales, it is extremely biased, and its noise budget is dominated by the CIB. When  SO LAT is combined with FYST, the CIB noise residual dominates much later $\ell \sim 4200$, but because the atmospheric noise is higher, there is no window over which the ILC-extracted tSZ power spectrum is not noise dominated. 

We note that the CIB residual noise reduction that FYST brings to SO LAT is not something that SO LAT could achieve on its own by simply having a longer survey time. This was tested by reducing the instrumental white noise in SO LAT bands by two, and we observed a reduction of the ILC residual instrumental noise in the reconstructed Compton-y map, but the other noise components remained the same. This ‘boosted’ version of SO LAT alone did not have a lower CIB residual power spectrum compared to the planned SO LAT or SO LAT plus FYST combination.

Because tSZ and CIB have a non-zero correlation in the Websky simulations and because FYST's high frequencies probe the CIB better, we used CILC to deproject the CIB component, modelling it with a simple one-component modified blackbody spectrum. Compared to the ILC case, adding FYST to SO LAT lowers the CIB noise residual by only $\sim 7\%$. The CMB and kSZ noise residuals are also reduced by $\sim 5\%$ when combining SO LAT and FYST when using CILC. However, the cumulative foregrounds are suppressed by $\sim 30\%$, and atmospheric noise is a bit lower for the SO LAT and FYST combination. Despite a lower reduction of the CIB residual noise when combining SO LAT and FYST, the CIB deprojection case offers a large  $\ell \in [1800,3500]$ window where the CILC-extracted Compton-y map is not noise dominated after instrumental noise and CMB debiasing. 

\begin{acknowledgements}
The authors would like to thank Kaustuv Basu for his guidance and advice throughout the project. Srinivasan Raghunathan, Jacques Delabrouille, and Frank Bertoldi for their time, valuable insights, and numerous helpful discussions. The referee for useful comments and numerous improvements to the draft. The editor Laura Pentericci for her advice and help on the submission process. Michael Coronado for his language editing work which significantly helped improve the manuscript. The CCAT-prime collaboration for their feedback on this work and Colin Hill for his help. Laila Linke for her useful comments on this draft. We furthermore acknowledge support from the International Max-Planck Research School (IMPRS) and the Bonn-Cologne Graduate School of Physics and Astronomy (BCGS). The simulations of the CIB used in this paper were developed by the Websky Extragalactic CMB Mocks team, with the continuous support of the Canadian Institute for Theoretical Astrophysics (CITA), the Canadian Institute for Advanced Research (CIFAR), and the Natural Sciences and Engineering Council of Canada (NSERC), and were generated on the GPC supercomputer at the SciNet HPC Consortium. SciNet is funded by the Canada Foundation for Innovation under the auspices of Compute Canada, the Government of Ontario, Ontario Research Fund – Research Excellence, and the University of Toronto.
\end{acknowledgements}

% for the bibliography, at the end

%\input{old.bbl}
%\bibliography{sample}
%\bibliographystyle{aa}

\begin{appendix}

\section{\textit{Planck} data} \label{AnnexeP}

\subsection{Internal linear combination and constrained internal linear combination weights}  \label{AnnexeP_w}

We applied our pipeline to \textit{Planck} High-Frequency instrument (HFI) Data Release 2\footnote{\url{https://irsa.ipac.caltech.edu/data/Planck/release_2/all-sky-maps/}}  \citep[DR2;][]{Planck_2016d} in order to see how well we could recover the true $y$ using an ILC. We also considered a CILC to deproject CIB by constraining the weights so that they have a null response to the frequency spectrum of the CIB (see Eq. \eqref{eqn:CILC_w}). However, the CIB deprojection is not perfect, because the CIB frequency spectrum depends on the population of the sources and their redshift. 

Looking at the ILC weights gave us information about which channel contributes the most to extracting the tSZ signal and minimising the noise. These were the two conditions that the weights had to satisfy (see Eq. \eqref{eqn:ILC_cond1} and \eqref{eqn:ILC_cond2}).
The ILC weights ponderated by their respective frequency mixing vector value showed the fractional contribution of each frequency channel to the recovered y. The fractional contribution of the $\sim 220$GHz band was null because the value of the mixing vector at this frequency is null. 

\begin{figure}[h!]
    \centering
    \includegraphics[width=0.48\textwidth]{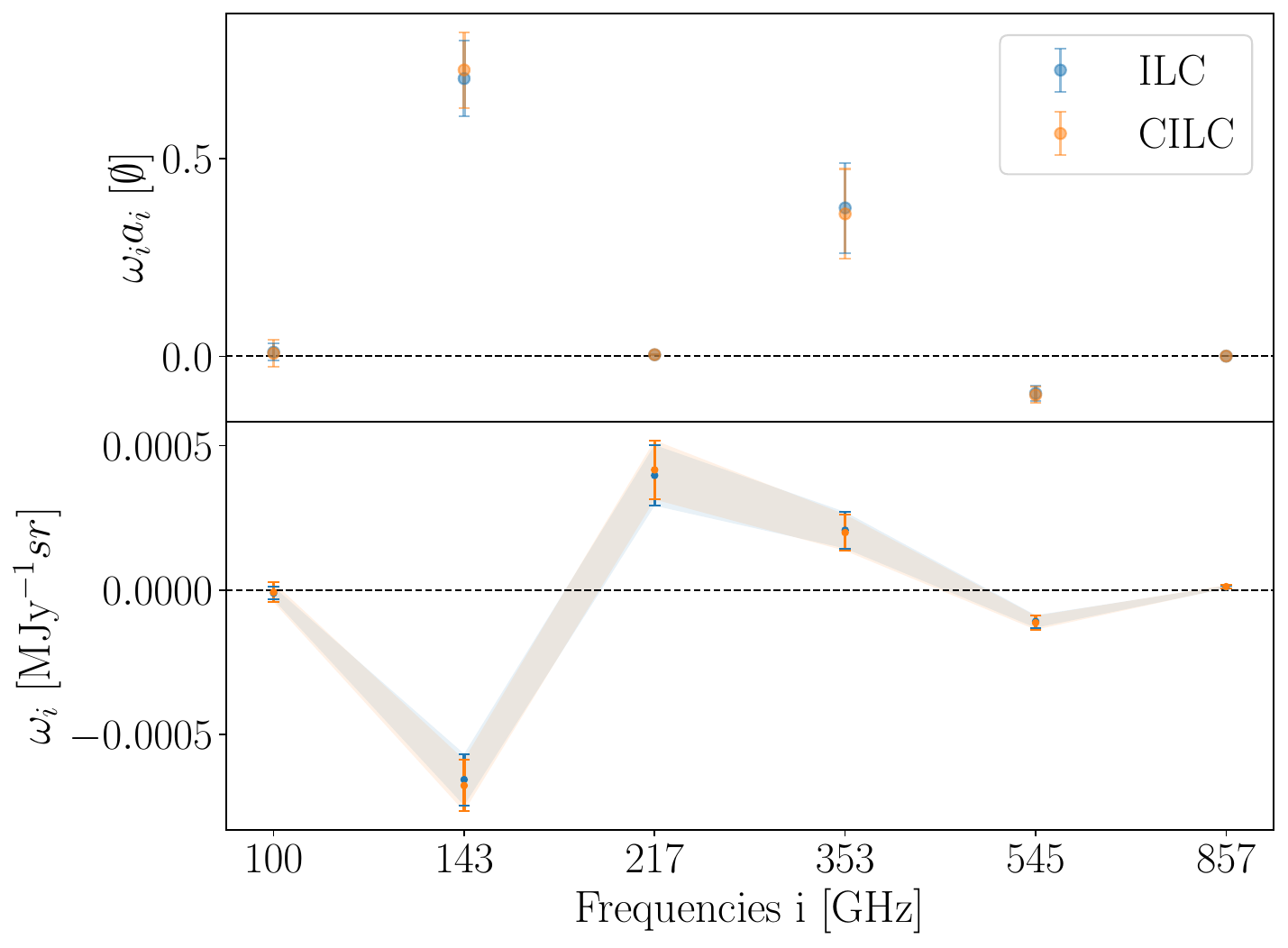}
    \caption{Internal linear combination (blue) and CILC (orange) weights with CIB deprojected on \textit{Planck} data. Top figure: Fractional contribution of each frequency channel to the recovered $y$-map. Bottom figure: Frequency channels that contribute to maximising the tSZ signal and minimising the noise. The error bars show the standard deviation around the mean weight of all of the $192$ tessellated fields for each frequency map.}
    \label{fig:w_Planck}
\end{figure}

For a map-based ILC applied to \textit{Planck} data (see blue points in Fig. \ref{fig:w_Planck}), we concluded from the weight that the 143 GHz channel is the one that contributes the most, with a factor of $\omega _{143}a_{143}\approx 0.7$, to the reconstructed final map as well as to minimising the noise. The 217 GHz channel does not contribute to the final reconstructed $y$ map because the tSZ signal is null at this frequency. However, its high $\omega _{217}$ value makes it the main contributor to noise minimisation. This was expected, as the tSZ signal is absent at this frequency and the map contains all the noise contaminants. Next are the contributions of the $353$ GHz channel and the $545$ GHz channel. The contribution values of those channels are consistent with previous studies \citep{Jens_2018}. In the case where a CILC is applied to \textit{Planck} data to deproject the CIB (orange points, Fig. \ref{fig:w_Planck}), we saw that the weights are almost identical to the ILC ones. This is expected, as \cite{Alonso_2018} also found that deprojecting CIB with a CILC does not result in much less CIB contamination in the reconstructed tSZ profile, arguing that this shows the ILC already does a sufficient job of cleaning the CIB.   

\subsection{Processing of the maps}

\begin{figure}[h!]
    \centering
    \includegraphics[width=0.48\textwidth]{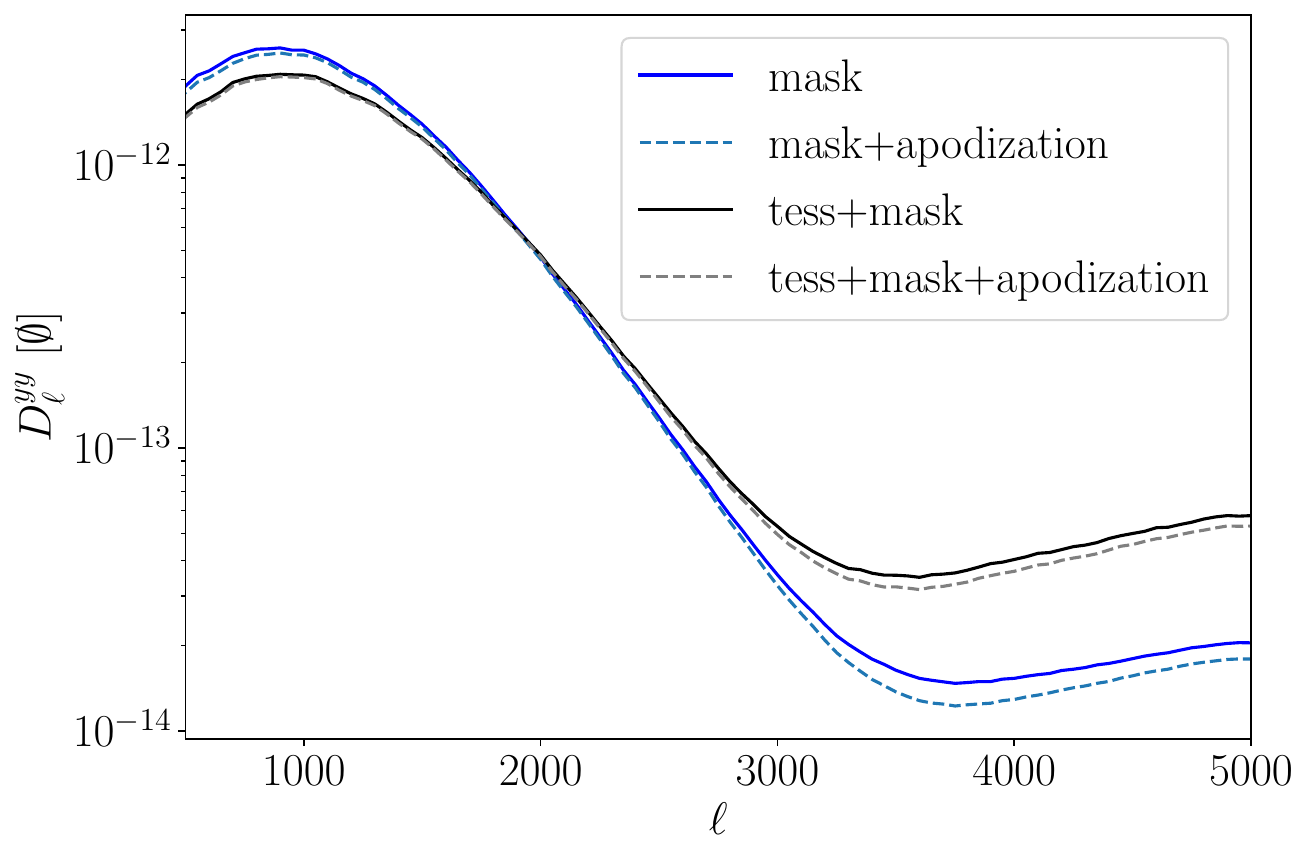}
    \caption{Power spectra of the Compton-y map using our ILC pipeline on \textit{Planck} DR2 with different configurations. In blue is the power spectrum of the extracted map with a simple $f_{sky}=0.6$ Galactic mask. In black is the same thing but with a tessellation of the sky. The dotted lines (blue and black) indicate the inclusion of apodisation in each case.}
    \label{fig:tess_mask_apo}
\end{figure}

We tested the consistency and effect of each of the post-processing steps applied on the map (see Fig. \ref{fig:tess_mask_apo}). The blue curve in the figure shows the ILC-extracted power spectrum computed with a Galactic mask and without tessellation (see Fig. \ref{fig:masks}), while the lighter blue dotted curve shows the same but  with apodisation. The black curve shows the case with the tessellation of the sky and the ILC being applied to each rhombus separately. The grey dotted line shows when apodisation was also applied. For our particular case of tSZ extraction, the mask apodisation only lowers the power spectra of the maps by less than $5\%$. Comparing the power spectrum with tessellation (black) and without (blue) showed that tessellation helps reduce the noise on large scales, up to $\ell \approx 2000$, by a factor of approximately two. At $\ell = 2500$, it increases the noise by around 15\%. This can be understood as the size of the rhombus being small enough to break down regions highly contaminated by unwanted astrophysical components on large scales and to weigh down such regions in order to better recover the tSZ signal. But the rhombus size might be too big to optimally deal with smaller region noise. Moreover, with 192 total rhombuses over the sky, we might be creating border effects on smaller scales and thus increasing the noise residual on those scales. The scales ($\ell > 2200$) are also affected by the $\sim 9.66$' beam.

\section{Constrained ILC: Deprojecting the cosmic microwave background}\label{AnnexeCILC}

\begin{figure}[h!]
    \centering
    \includegraphics[width=0.48\textwidth]{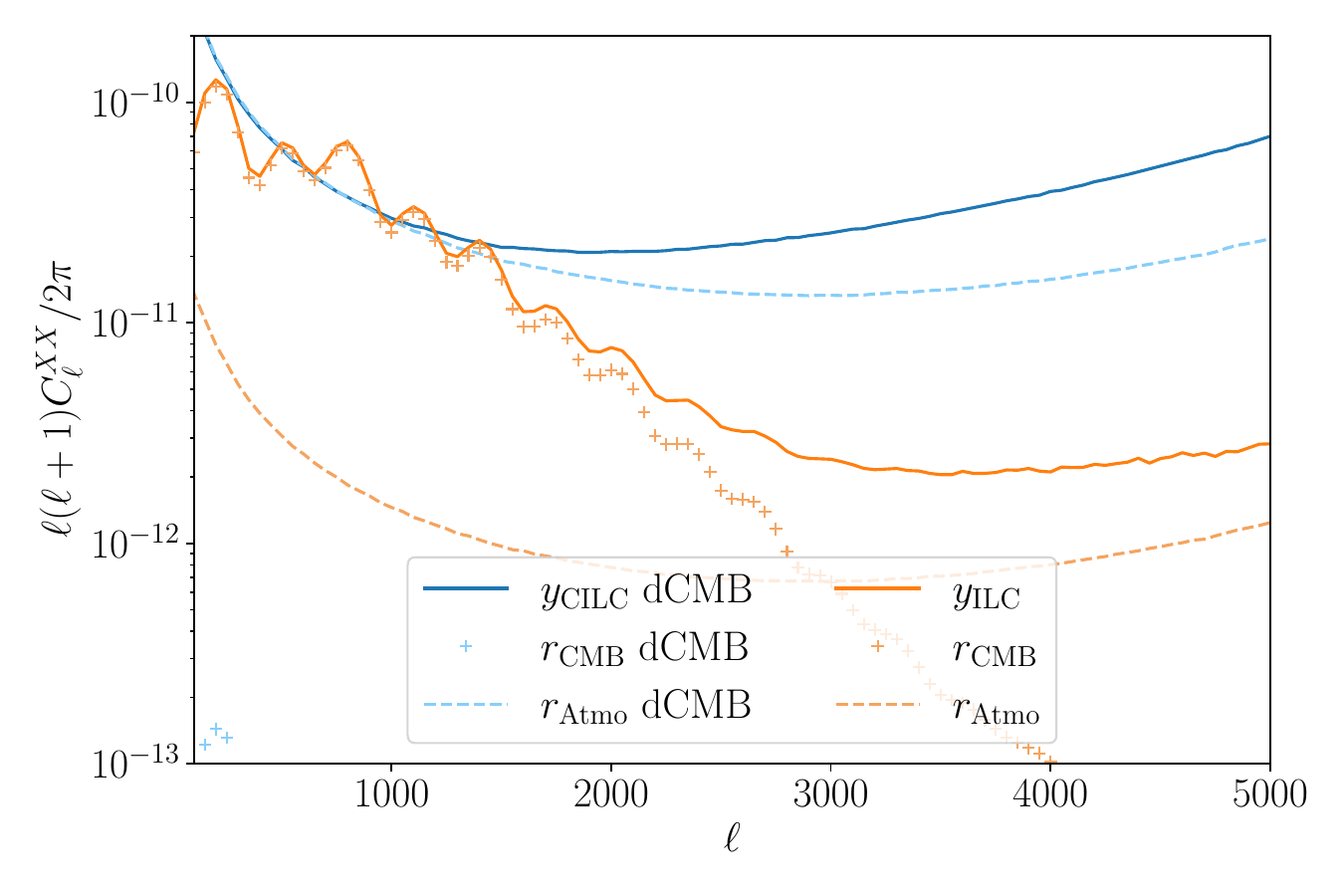}
    \caption{Power spectra of ILC/CILC (full lines) and their noise residuals (dotted lines). Specifically, the orange curve is the power spectrum of the ILC-extracted Compton-$y$ map. The lighter orange lines are the noise residuals. Specifically, they are the ILC residual CMB noise ($r_{\mathrm{CMB}}$) and  $r_{\mathrm{Atmo}}$ for the atmospheric noise. The blue lines represent the case with CILC where the  CMB is deprojected. The solid, darker blue curve is the power spectrum of the CILC-extracted Compton-$y$ map. The lighter blue curves are the noise residuals.}
    \label{fig:dCMBvsNone}
\end{figure}

When applying an ILC to extract back a Compton-$y$ map, the CMB is the dominant noise residual ($r_{\mathrm{CMB}}$) on large scales (see Fig.\ref{fig:dCMBvsNone}). Using CILC to deproject the CMB, the residual CMB noise is suppressed by three orders of magnitude. However, the additional constraints imposed on the weights to deproject the CMB led to a smaller solution space and therefore increased noise. This led to a one order of magnitude increase of the atmospheric residual noise ($r_{\mathrm{Atmo}}$), which worsens the estimate of the extracted Compton-$y$ power spectrum, especially on small scales, by more than one order of magnitude. This is commented on extensively in Sec. \ref{CILC_results} and was tested using our Skymodel pipeline (see Sec. \ref{des_skymodel}).

\section{ILC weights} \label{Annexew}

\begin{figure}[h!]
    \centering
    \includegraphics[width=0.50\textwidth]{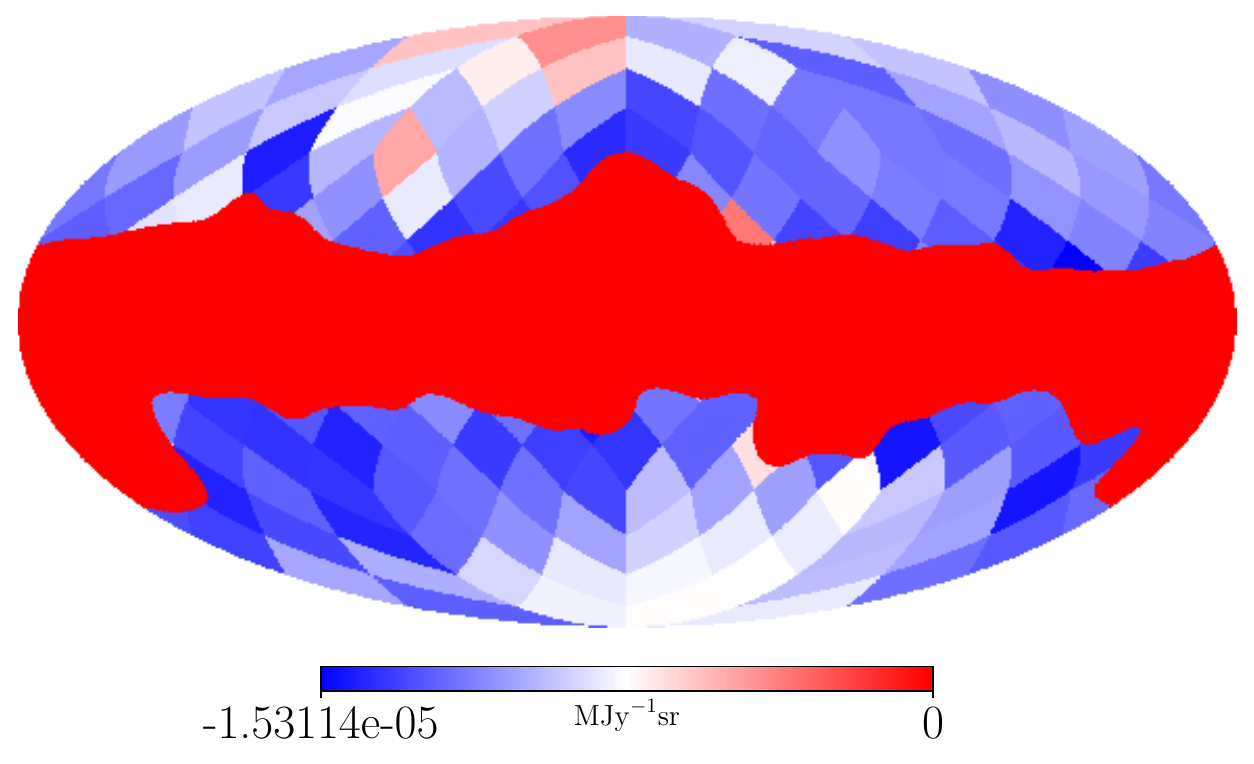}
    \caption{Maps of the weights for each patch of the sky where the ILC was applied for the 860 GHz frequency band. The weights were computed from simulations that included all Galactic foregrounds and extragalactic components (see Table \ref{tab:skymodel}) at the sensitivities, beam sizes, and frequencies of SO LAT and FYST (see Table \ref{tab:fsb}) and with 90\% correlated atmospheric red noise generated using the noise curves given by \cite{Choi_2020_Noise} (see Fig. \ref{fig:red_noise}).}
    \label{fig:maps_weights}
\end{figure}

In our study, we tesselated the sky into 192 rhombi of equal area using the HEALPix nested scheme. This sky separation allowed us to better
adapt to the spatially dependant contaminant of some Astrophysical emissions, such as the Galactic foregrounds. The ILC we applied on each zone resulted in one weight per frequency and per rhombus. An example of a full sky weights map produced and used by our ILC can be seen in Fig. \ref{fig:maps_weights}. 

\section{Simulated maps} \label{AnnexeSky}

\begin{figure*}[h!]
    \centering
    \includegraphics[width=0.35\textwidth]{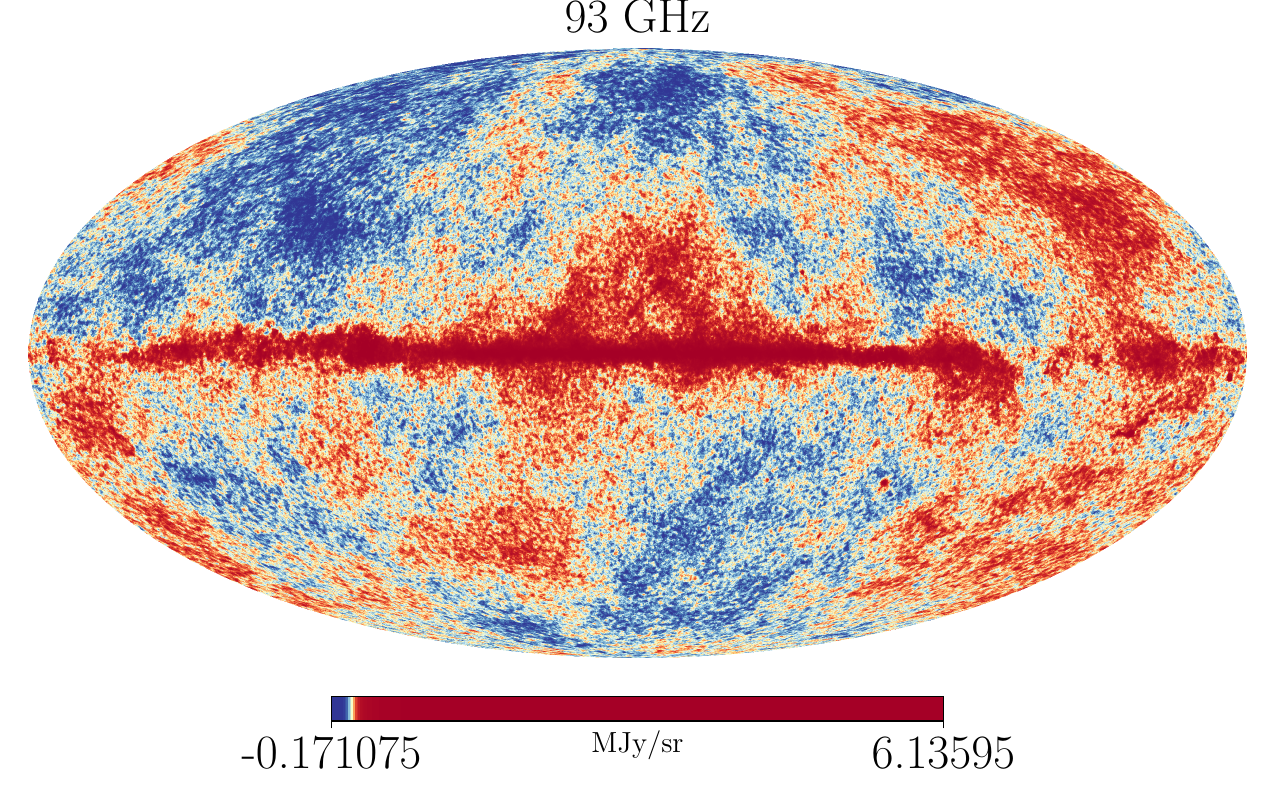}\includegraphics[width=0.35\textwidth]{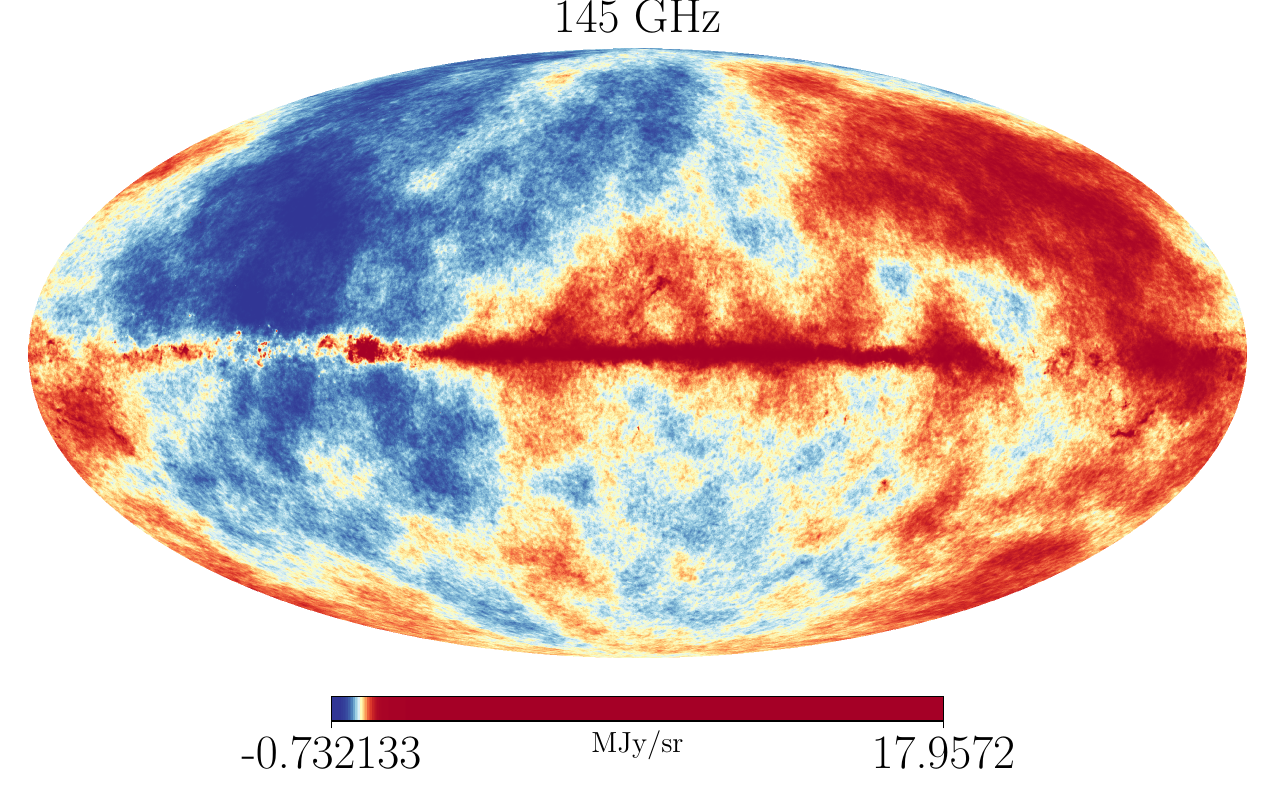}
    \includegraphics[width=0.35\textwidth]{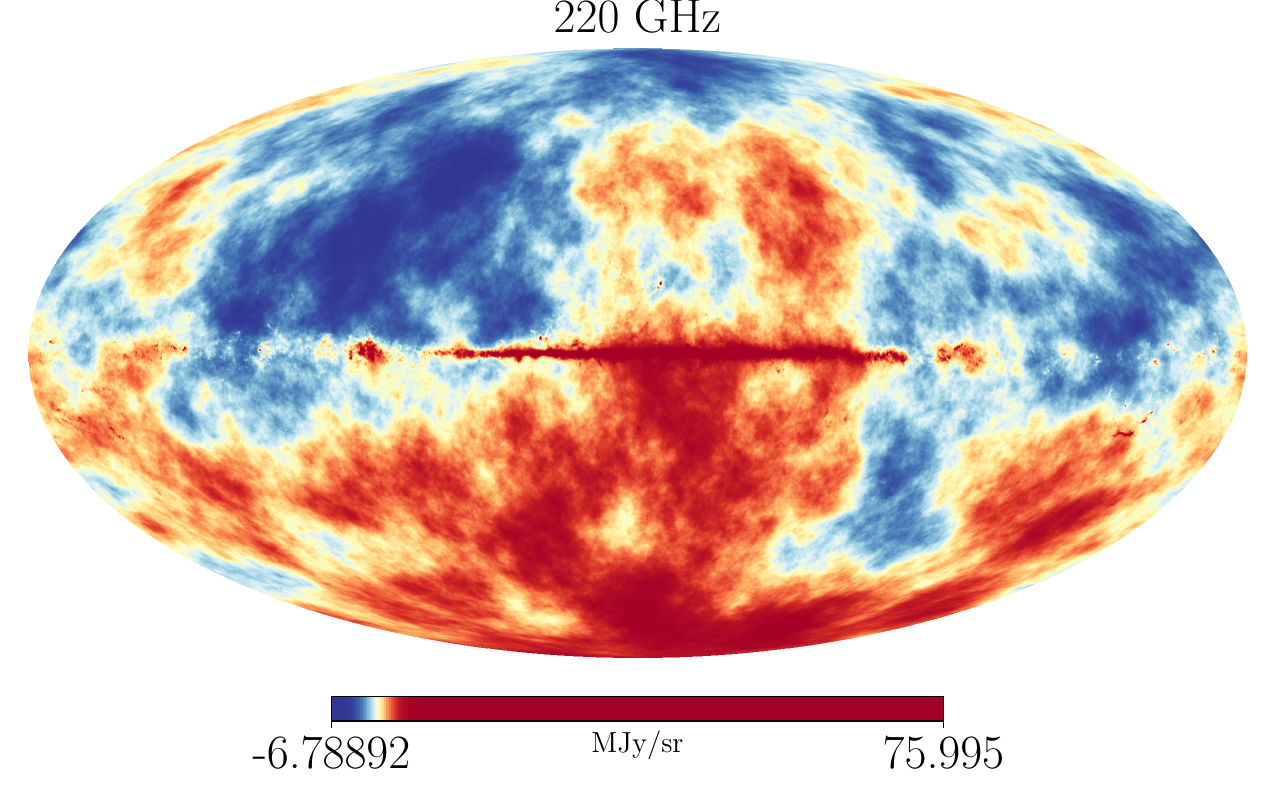}\includegraphics[width=0.35\textwidth]{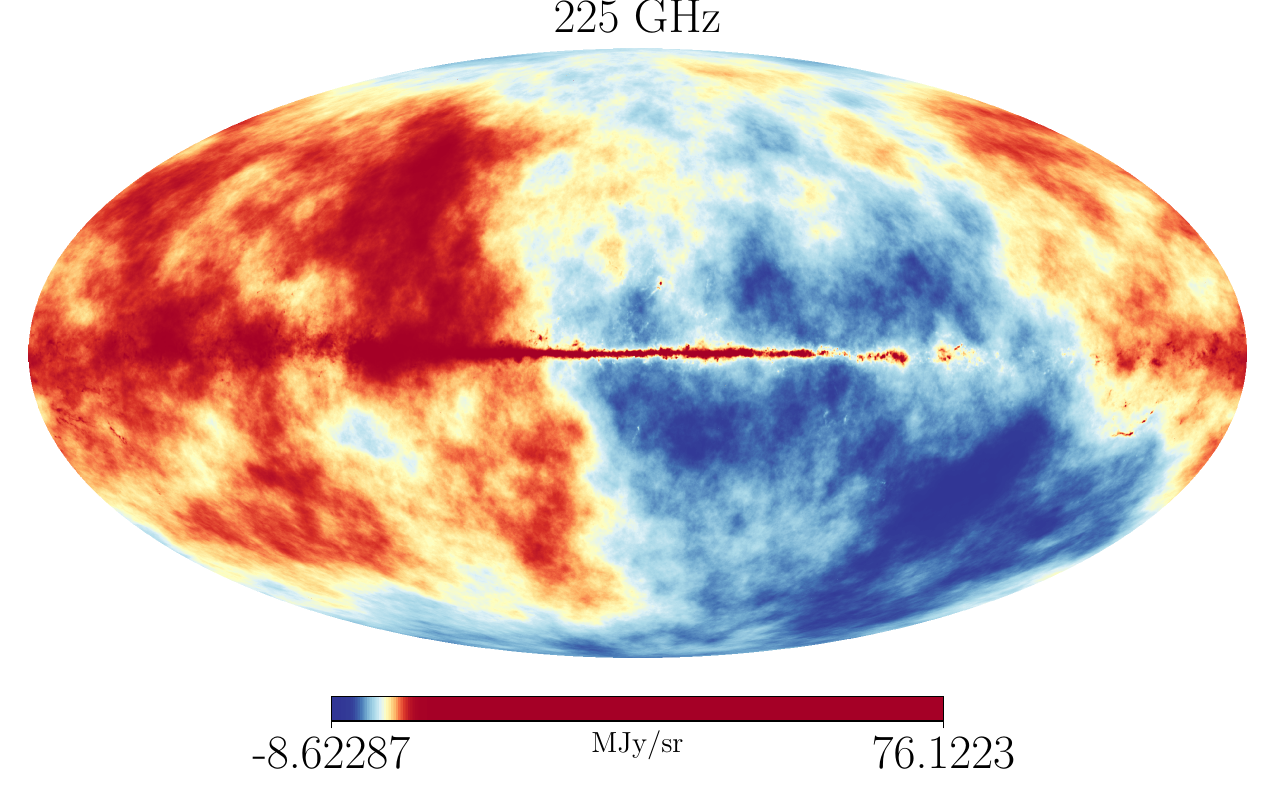}
    \includegraphics[width=0.35\textwidth]{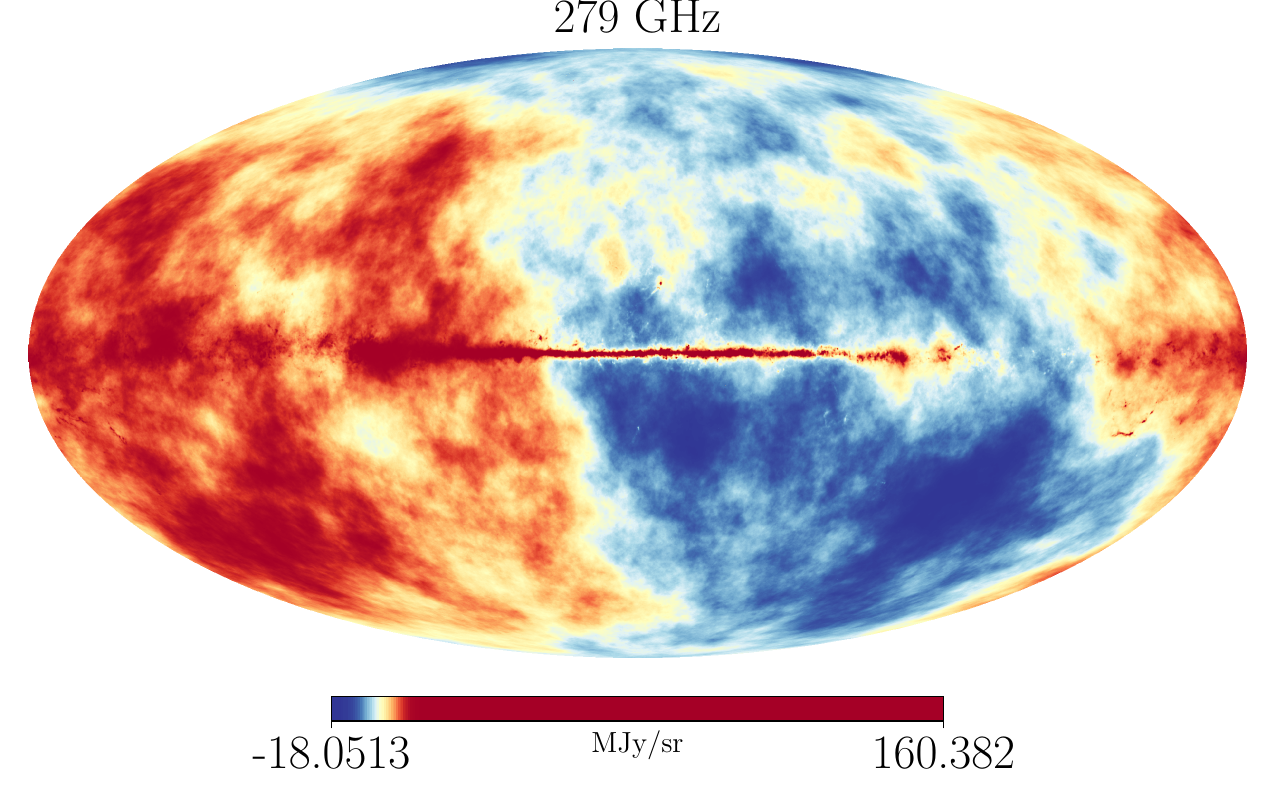}\includegraphics[width=0.35\textwidth]{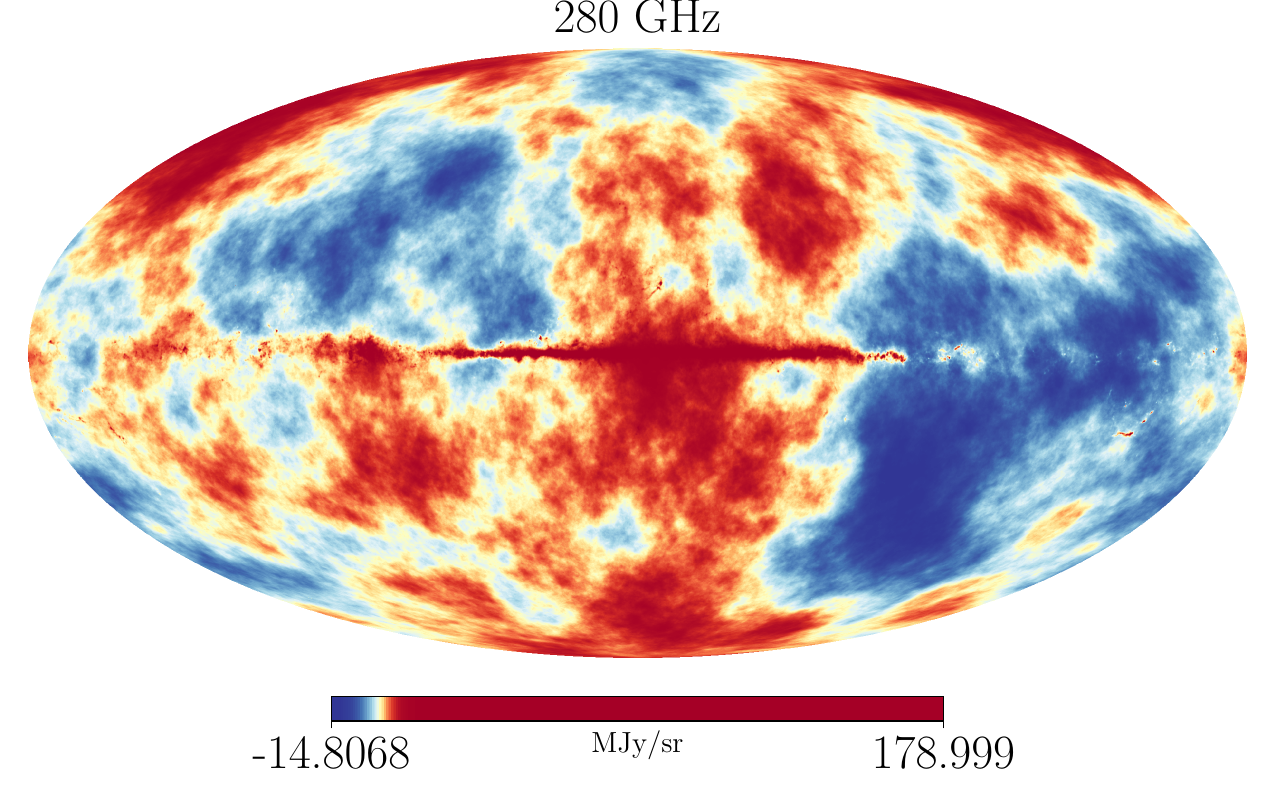}
    \includegraphics[width=0.35\textwidth]{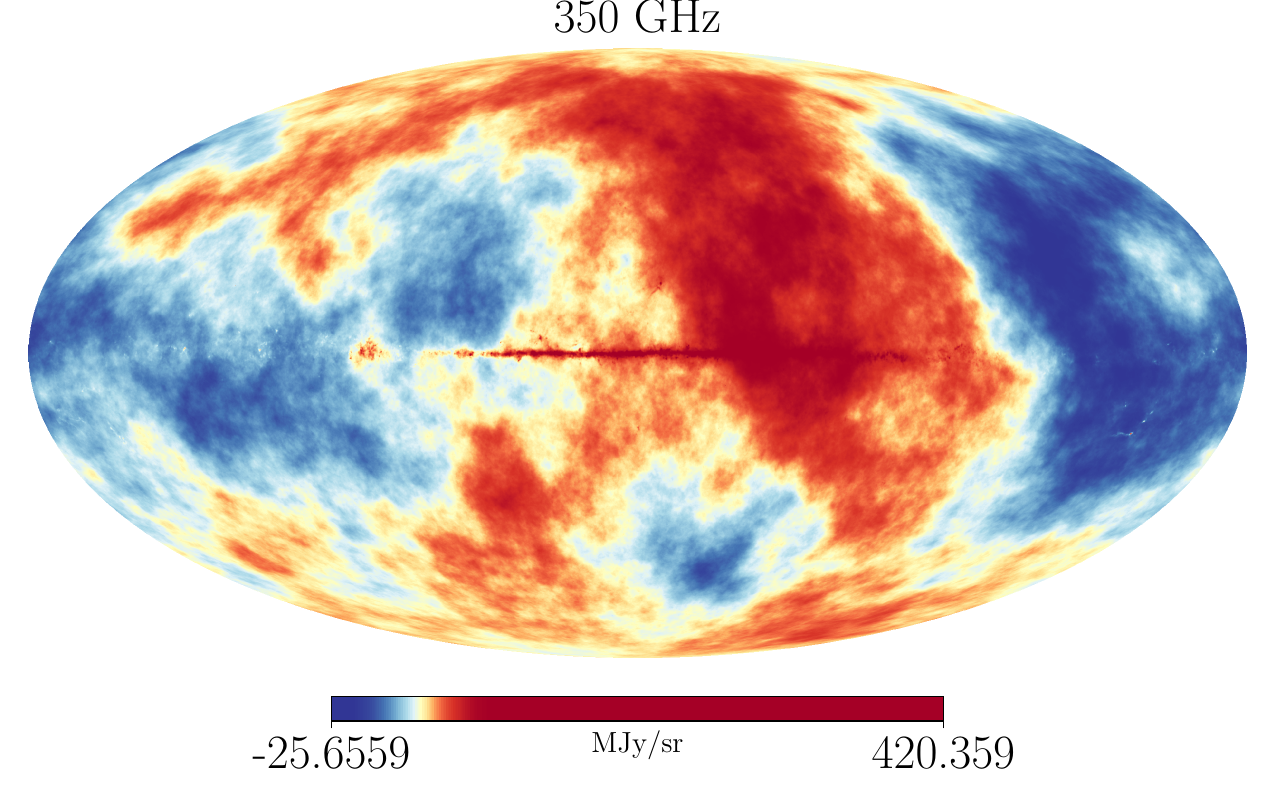}\includegraphics[width=0.35\textwidth]{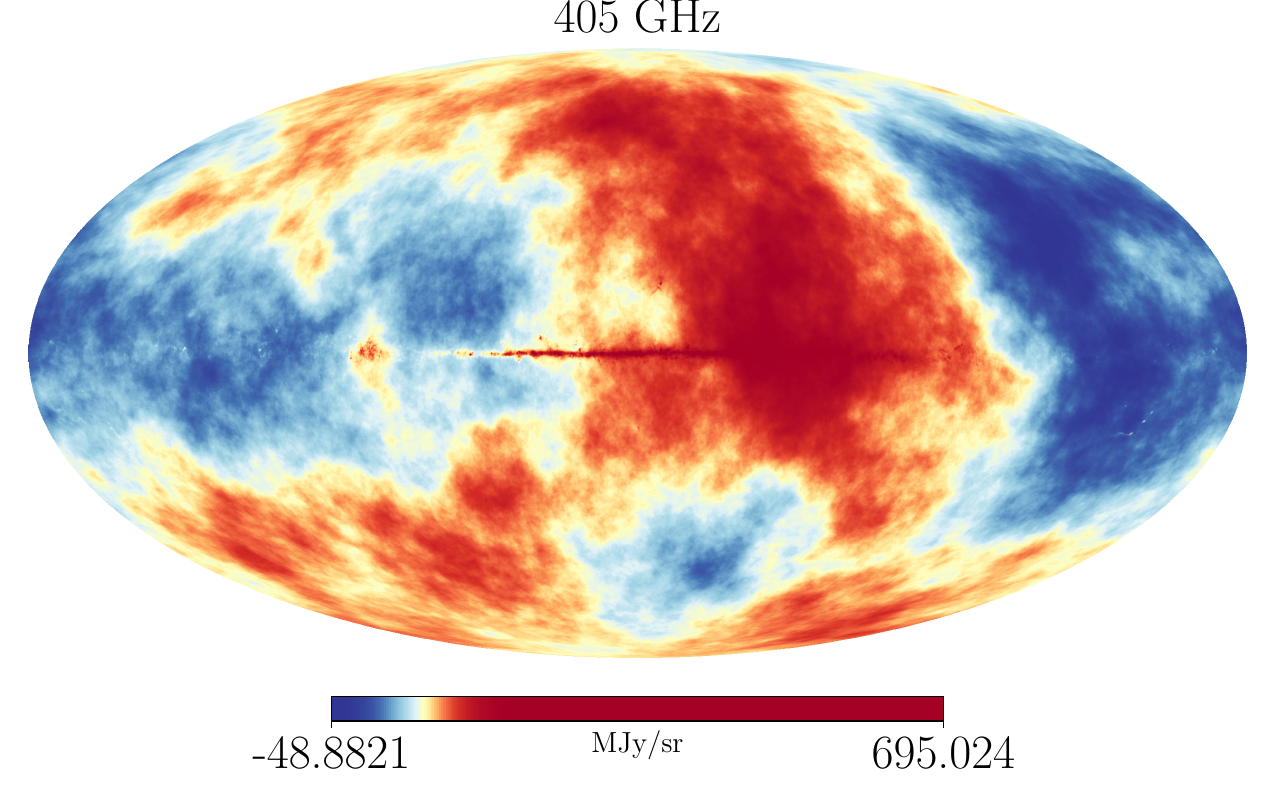}
    \includegraphics[width=0.35\textwidth]{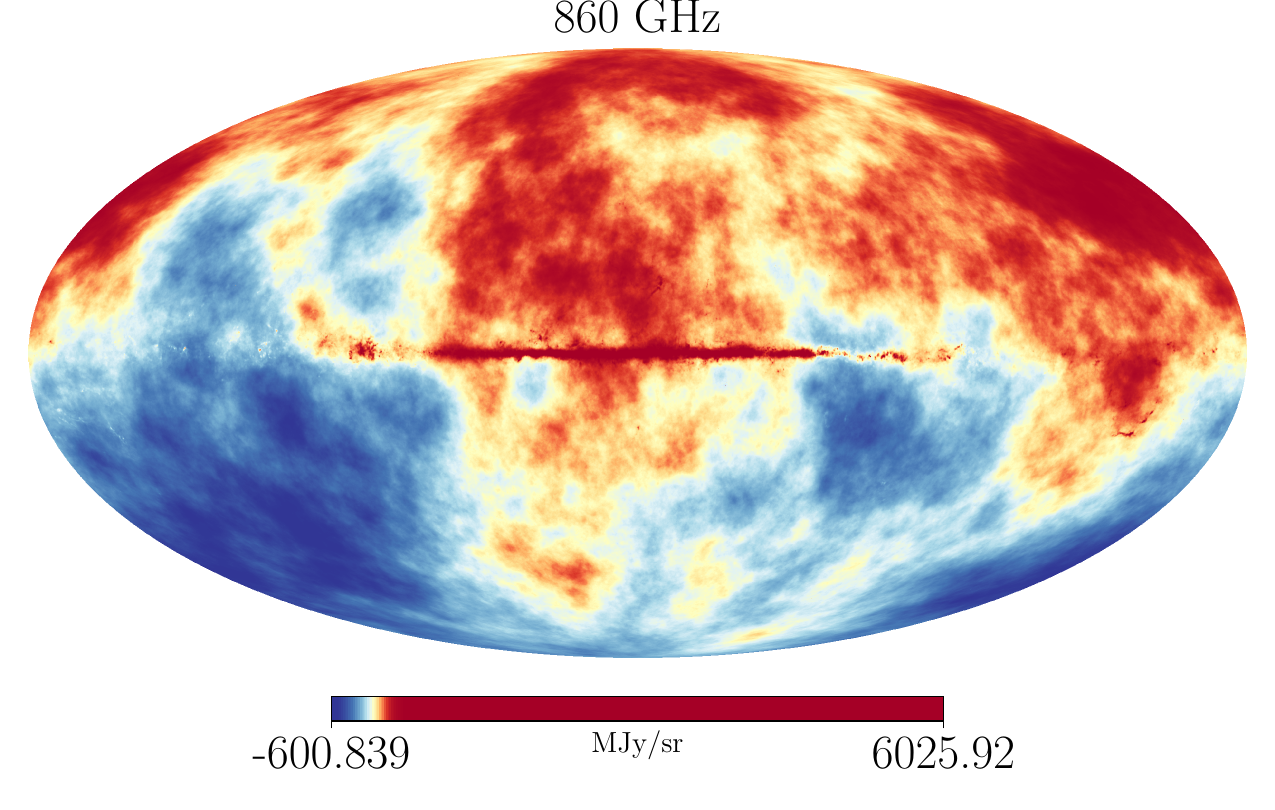}
    \caption{Mock HEALPy maps of the microwave sky simulated using the Skymodel pipeline (see Sec. \ref{des_skymodel}). All Galactic foregrounds and extragalactic components (see Table \ref{tab:skymodel}) at the sensitivities, beam sizes, and frequencies of SO LAT and FYST (see Table \ref{tab:fsb}) and the 90\% correlated atmospheric red noise between neighbouring channels generated using the noise curves given by \cite{Choi_2020_Noise} were included (see Fig. \ref{fig:red_noise}).}
    \label{fig:mock-maps-sky}
\end{figure*}

In this work, we used the Skymodel pipeline based on PySM and the Websky simulations to generate mock maps of the microwave sky at the frequencies, sensitivities, and instrument beam size of the upcoming SO LAT and FYST (see Table \ref{tab:fsb}). The atmospheric red noise was generated using the noise curves of \cite{Choi_2020_Noise} and the red noise of the neighbouring frequency channels of a given telescope were correlated at 90\% using Eq. \ref{eqn:corr_noise}. The resulting simulated microwave sky maps used can be seen in Fig. \ref{fig:mock-maps-sky}. 

%\twocolumn

\end{appendix}

\end{document}